\documentclass[a4paper,12pt]{article}
\usepackage[utf8]{inputenc}      
\usepackage{hyperref}
\usepackage{braket}
\usepackage{bm}

\usepackage{longtable} 
\usepackage{multirow}
\usepackage{booktabs}
\usepackage{siunitx}

\usepackage{mathtools}
\usepackage{amsmath}
\usepackage{amssymb}
\usepackage{amsfonts}
\usepackage{amsthm}
\usepackage{setspace}
\usepackage{pgfplots}
\usepackage{graphicx}
\usepackage{ifthen}
\usepackage{amscd}
\usepackage{color}
\usepackage{mathbbol}
\usepackage{url}     
\usepackage[a4paper, total={6in, 9.5in}]{geometry}
\usepackage{subcaption}
\usepackage{bbold}
\usepackage{enumerate}

\newtheorem{thm}{Theorem}[section] 
\newtheorem{ass}{Assumption}

\newtheorem{prop}[thm]{Proposition} 
\theoremstyle{definition}

\theoremstyle{remark}

\numberwithin{equation}{section}

\numberwithin{equation}{section}

\usepackage{authblk}
\usepackage[authoryear]{natbib}

\author{Giuseppe Buccheri\footnote{Department of Economics, University of Verona, Verona, Italy. Email: giuseppe.buccheri@univr.it}, Fulvio Corsi\footnote{Department of Economics and Management, University of Pisa. Email: fulvio.corsi@unipi.it}, and Emilija Dzuverovic\footnote{Dep. of Economics and Management, University of Pisa. Email: emilija.dzuverovic@ec.unipi.it} }

\begin{document}

\setlength{\abovedisplayskip}{10pt}
\setlength{\belowdisplayskip}{10pt}

\newgeometry{left=2.5cm,right=2.5cm,top=1cm, bottom=2cm}

\title{From rotational to scalar invariance:\ Enhancing identifiability in score-driven factor models\footnote{This work was supported by the Italian Ministry of University and Research under the PRIN projects \textit{Dynamic models for a fast changing world:\ An observation-driven approach to time varying parameters} (grant n.\ 20205J2WZ4, CUP B37G22000180006), and \textit{Realized Random Graphs:\ A New Econometric Methodology for the Inference of Dynamic Networks} (grant n.\ 2022MRSYB7, CUP B53D23010090006).} }
\vspace{0.5cm}
\date{}
\maketitle
\vspace{-0.8cm}
\begin{center}
November, 2024\\ 
\vspace{0.5cm}
\end{center}

\begin{abstract} 

We show that, for a certain class of scaling matrices including the commonly used inverse square-root of the conditional Fisher Information, score-driven factor models are identifiable up to a multiplicative scalar constant under very mild restrictions.\ This result has no analogue in parameter-driven models, as it exploits the different structure of the score-driven factor dynamics.\ Consequently, score-driven models offer a clear advantage in terms of economic interpretability compared to parameter-driven factor models, which are identifiable only up to orthogonal transformations.\ Our restrictions are order-invariant and can be generalized to score-driven factor models with dynamic loadings and nonlinear factor models.
We test extensively the identification strategy using simulated and real data.\ The empirical analysis on financial and macroeconomic data reveals a substantial increase of log-likelihood ratios and significantly improved  out-of-sample forecast performance when switching from the classical restrictions adopted in the literature to our more flexible specifications.

\vspace{0.2cm}

\vspace{1cm}

\end{abstract}

\newpage

\newgeometry{left=2.5cm,right=2.5cm,top=2cm, bottom=2cm}

\section{Introduction}

Given the increasing interconnectedness of the global economy, the dynamics of a panel of economic and financial time series can often be described by a set of few common factors.\ For this reason, the use of dynamic factor models has become increasingly popular in the econometric literature.\ In a dynamic factor model, the common factors may be either observable or latent.\ Assuming latent factors is regarded as less restrictive because it sidesteps the challenge  of selecting the common factors from a vast array of potential explanatory variables.\ 
On the other hand, the use of unobservable factors generates identification issues, the solution of which poses significant interpretability problems.\ For instance, the estimated factors might be order-dependent or defined up to rotations, thereby complicating their economic or financial interpretation.

In this paper, we contribute to the literature on factor models identification. In particular, we focus on factor models where the factor dynamics are driven by the score of the conditional observation density.\ Using the language of \cite{Cox}, these models are called \textit{observation-driven}, meaning that the main source of variation of the factor dynamics are the past observations.\ In contrast, in parameter-driven models, the factor dynamics depend on their own source of uncertainty; see, e.g., \cite{geweke1977dynamic}, \cite{chamberlain1983funds}, \cite{StockWatson2011}, \cite{doz2012quasi}.\ Examples of score-driven factor models are given in \cite{creal2014observation}, \cite{oh2018time}, \cite{artemova}, among others.\ Score-driven factor models are particular instances of the general class of score-driven models proposed by \cite{harvey2013dynamic} and \cite{creal2013generalized}.\ The main advantage of score-driven factor models with respect to parameter-driven specifications is that the exact likelihood can always be written in closed form.\ Moreover, the law of motion of the factors, being driven by the score of a non-normal conditional density, is naturally robust to the extreme observations which may occur during periods of financial and/or economic distress.

The main contribution of this paper is to show that score-driven factor models can be identified under significantly milder restrictions compared to parameter-driven factor models.\ This is due to the fact that, when subject to a linear non-singular  transformation, the law of motion of the factors in a score-driven model is invariant only under two specific choices of the scaling matrix used to normalize the score.\ The first choice corresponds to the inverse of the conditional Fisher information matrix, while the second one corresponds to the identity matrix.\ In these two cases, multiplying the factors by a non-singular matrix leads to new factors following a score-driven recursion of the same form.\ Therefore, both the static and time-varying parameters are not identifiable, as in parameter-driven factor models.\ However, when the scaling matrix is set differently, for example as the commonly used square-root inverse of the conditional Fisher information, multiplying the factors by a non-singular matrix results in a new recursion that cannot be represented as the original score-driven recursion.\ In other words, under affine non-singular transformations, the original score-driven law of motion is not generally preserved.\ Thus, for this class of normalization matrices, score-driven factor models behave differently compared to parameter-driven factor models, and the lack of invariance can be exploited to identify the static parameters.  

When the score is scaled through the class of normalization matrices for which the score-driven dynamics are not preserved, the factors and the loading matrix are identifiable up to a common multiplicative constant.\ This facilitates enormously their economic interpretability because a scalar transformation only scales the factors without altering their dynamics.\ Similarly, it leaves the relative magnitudes of the factor loadings unchanged, so that it is simple to quantify the impact each factor has on each variable.\ By comparison, parameter-driven factor models are identifiable only up to orthogonal transformations, and any identifying restriction amounts to selecting a specific rotation of the factors.\
Another implication of our results is that the estimated factors are order-invariant, meaning that they are independent from the order of the variables within the panel of time-series. This is an important advantage compared to other identification approaches in likelihood-based methods imposing \textit{a priori} dependencies structures between the factors and the observable data relying on the ordering of the time-series.\ 

A second contribution of the paper is to show that the same identification scheme can be used in score-driven factor models with time-varying loadings and in nonlinear factor models.\ Empirical evidence of time-varying loadings has been found in macroeconomic data (\citealt{mikkelsen2019consistent}, \citealt{xu2022testing}, \citealt{hillebrand2023exchange}) and financial data (\citealt{ adrian2009learning},  \citealt{kelly2019characteristics}, \citealt{giglio2022factor}).\ Nonlinear factor models with score-driven dynamics are widely adopted to reduce dimensionality in multivariate models.\ One example is given by the dynamic correlation model proposed in \cite{creal2011dynamic}, where the authors suggest to impose a factor structure for the time-varying parameters in order to reduce the size of the parameter space.

A relevant implication of the scalar invariance of score-driven factor models is that the static and dynamic parameters can be identified without imposing constraints on the factor loadings.\ The main identifying assumption needed is the diagonal structure of the matrix multiplying the scaled score, which is a standard restriction adopted in score-driven models.\ This leads to a higher flexibility in the specification of the model compared to parameter-driven factor models, where the factor loadings are generally constrained in order to guarantee identification.\ The higher flexibility is shown in our empirical application, where we find a significant increase in likelihood ratios when moving from the classical restrictions adopted in the literature to our specification.\ We also show that the factors extracted through our approach have a straightforward economic interpretation and produce superior out-of-sample forecasts of macroeconomic variables.


Another work addressing the problem of identifying the static and time-varying parameters in score-driven factor models is \cite{artemova}.\ The main difference between our approach and theirs is that \cite{artemova} normalizes the score through the inverse conditional Fisher information.\ This normalization preserves the structure of the score-driven law of motion when the time-varying parameters are multiplied by a non-singular matrix, leading to the same rotational indeterminacy found in parameter-driven factor models.\ Indeed, the identifying restrictions used in \cite{artemova} coincide with the orthogonality conditions on the factor loading matrix suggested by \cite{bai2012statistical}, which pick a specific rotation of the factors.\ In contrast, by normalizing the score through the inverse square-root of the conditional Fisher information, we exploit the different transformation properties of the score-driven law of motion in order to circumvent rotational indeterminacy and identify the model up to a common scalar constant.

The rest of the paper is organized as follows.\ In Section \eqref{ident:static}, we discuss the main differences between parameter-driven and score-driven factor models when it comes to the identification of the static and dynamic parameters.\ Section \eqref{sec:scalIdent} illustrates the main result of the paper, i.e., it provides a set of restrictions enabling the identification of score-driven factor models up to a common scalar constant.\ In Section \eqref{sec:gener}, we show that the same restrictions are sufficient to identify the model in the presence of time-varying loadings and in nonlinear factor structures.\ Several Monte Carlo results supporting  our conclusions are reported in Section \eqref{sec:MC}, while Section \eqref{sec:EA} presents the empirical results.\ Finally, Section \eqref{sec:concl} concludes.\ The proofs of the main results are collected in the Appendix section.

\section{Identification in parameter-driven \textit{versus} score-driven factor models}
\label{ident:static}

We start by illustrating the different behavior of parameter-driven and score-driven models when scaling the factors through a non-singular square matrix.\ Let us first consider the following parameter-driven factor model
\begin{align}
\bm{y}_t &= \bm{\Lambda}\bm{\alpha}_t + \bm{\epsilon}_t\label{eq:dfm:obs}\\
\bm{\alpha}_{t+1} &= \bm{\omega} +\bm{\Phi}\bm{\alpha}_t + \bm{\eta}_t\label{eq:dfm:trans}
\end{align}
where $\bm{y}_t$ is an $n\times 1$ vector, $\bm{\alpha}_t$ is a vector of $r<n$ unobservable common factors and $\bm{\Lambda}$ is an $n\times r$ matrix of factor loadings.\ The two white noises $\{\bm{\epsilon}_t\}_{t\in\mathbb{Z}}$, $\{\bm{\eta}_t\}_{t\in\mathbb{Z}}$ are independent, with covariance matrices $\bm{H}\in\mathbb{R}^{n\times n}$, $\bm{Q}\in\mathbb{R}^{r\times r}$, respectively.\ Let $\bm{T}\in\mathbb{R}^{r\times r}$ be a non-singular square matrix.\ The above model can be re-parameterized as follows
\begin{align}
\bm{y}_t &= \bm{\Lambda}^{*}\bm{\alpha}_t^* + \bm{\epsilon}_t\\
\bm{\alpha}_{t+1}^{*} &= \bm{\omega}^{*} + \bm{\Phi}^{*}\bm{\alpha}_t^{*} + \bm{\eta}_t^{*}
\end{align}
where $\bm{\Lambda}^{*}=\bm{\Lambda}\bm{T}$, $\bm{\alpha}_t^*=\bm{T}^{-1}\bm{\alpha}_t$, $\bm{\omega}^* = \bm{T}^{-1}\bm{\omega}$, $\bm{\Phi}^{*}=\bm{T}^{-1}\bm{\Phi}\bm{T}$, while $\bm{\eta}_t^*$ has covariance $\bm{Q}^*=\bm{T}^{-1}\bm{Q}\bm{T}^{-1\prime}$.\ Since the two parameterizations are observationally equivalent, the model is not identifiable without prior restrictions.\ Specifically, we need $r^2$ restrictions, equal to the number of degrees of freedom of $\bm{T}$, in order to fully identify the model parameters.\ A common identification strategy consists in restricting the covariance matrix of $\bm{\alpha}_t$, or that of $\bm{\eta}_t$, to be the identity matrix.\ This restriction leaves the factors identifiable up to an orthogonal transformation.\ We thus need to fix the remaining $r(r-1)/2$ free parameters, i.e., we need to specify the particular rotation being estimated.\ This can be done by imposing some structure on the factor loading matrix.\ Several types of restrictions are possible, some of which are order-invariant, while others depend on the ordering of the variables; see, e.g., \cite{harvey_1990} and \cite{bai2012statistical}.\ 

Let us now consider the identification of the static parameters in score-driven factor models.\ Let $\bm{\mathcal{F}}_{t}=\sigma(\bm{y}_t,\bm{y}_{t-1},\dots,\bm{y}_1)$ denote the information filtration generated by the process $\{\bm{y}_t\}_{t\in\mathbb{N}}$.\ Let us consider a score-driven factor model of the following form
\begin{align}
\bm{y}_t &= \bm{\Lambda}\bm{f}_t + \bm{\epsilon}_t\label{eq:dfm:sd:obs}\\
\bm{f}_{t+1} &= \bm{c} +\bm{A}\bm{s}_t + \bm{B}\bm{f}_t\label{eq:dfm:sd:trans}
\end{align}
where $\bm{\epsilon}_t$ is a white noise with covariance $\bm{\Sigma}\in\mathbb{R}^{n\times n}$, $\bm{s}_t = \bm{S}_t \bm{\nabla}_t$, and
\begin{equation}
\bm{\nabla}_t = \left[\frac{\partial\log p(\bm{y}_t|\bm{\mathcal{F}}_{t-1},\bm{f}_t)}{\partial\bm{f}_t}\right]'\label{eq:score}
\end{equation}
is the score of the predictive likelihood.\ The scaling matrix $\bm{S}_t$ is set as $\bm{S}_t =\left(\bm{\mathcal{I}}_{t|t-1}\right)^{-\beta}$, $\beta\in [0,1]$, where $\bm{\mathcal{I}}_{t|t-1}$ denotes the conditional Fisher information, defined as 
\begin{equation}
\bm{\mathcal{I}}_{t|t-1}=\mathbb{E}[\bm{\nabla}_t\bm{\nabla}_t'|\bm{\mathcal{F}}_{t-1}].
\end{equation}
Common choices of the exponent $\beta$ are $\beta=\left\{0,\frac{1}{2},1\right\}$.\ 
The time-varying parameter $\{\bm{f}_t\}_{t\in\mathbb{N}}$ is predictable with respect to the filtration $\bm{\mathcal{F}}_{t}$, implying that the predictive likelihood coincides with the conditional observation density $p(\bm{y}_t|\bm{f}_t)$ determined by the distribution of the idiosyncratic noise $\bm{\epsilon}_t$.\ The latter is left unspecified at the moment because the results we recover are independent from the choice of such a distribution.\ The score-driven recursion allows writing the likelihood in closed form via the prediction error decomposition, in a similar fashion to GARCH-type models, and to infer the static parameters through standard numerical optimization methods.\ Moreover, the update of the time-varying parameters based on the score provides robust estimates in the case of fat-tailed distributions; see also \cite{d2024dynamic}, \cite{opschoor2024conditional} and \cite{blasques2024maximum} for some recent applications of score-driven models.  

It is immediate to prove the following result.

\begin{prop}
Let $\bm{T}\in\mathbb{R}^{r\times r}$ be non-singular and let us set $\overline{\bm{\Lambda}}=\bm{\Lambda}\bm{T}$, $\overline{\bm{f}}_t=\bm{T}^{-1}\bm{f}_t$. The score-driven factor model in Equations \eqref{eq:dfm:sd:obs}, \eqref{eq:dfm:sd:trans} can be re-parameterized as follows:
\begin{align}
\bm{y}_t &= \overline{\bm{\Lambda}}\ \overline{\bm{f}}_t + \bm{\epsilon}_t\label{eq:dfm:sd:obsRep}\\
\overline{\bm{f}}_{t+1} &= \overline{\bm{c}} +\overline{\bm{A}}\  \left(\overline{\bm{\mathcal{I}}}_{t|t-1}\right)^{-\beta}(\bm{T}^{-1+\beta})^{\prime}\overline{\bm{\nabla}}_t    + \overline{\bm{B}}\ \overline{\bm{f}}_t\label{eq:dfm:sd:transRep}
\end{align}
where $\overline{\bm{c}}=\bm{T}^{-1}\bm{c}$, $\overline{\bm{A}}=\bm{T}^{-1}\bm{A}\bm{T}^{\beta}$, $\overline{\bm{B}}=\bm{T}^{-1}\bm{B}\bm{T}$, $\overline{\bm{\nabla}}_t = \left[\frac{\partial\log p(\bm{y}_t|\bm{\mathcal{F}}_{t-1},\overline{\bm{f}}_t)}{\partial\overline{\bm{f}}_t}\right]'$, $\overline{\bm{\mathcal{I}}}_{t|t-1}=\mathbb{E}[\overline{\bm{\nabla}}_t\overline{\bm{\nabla}}_t'|\bm{\mathcal{F}}_{t-1}]$.
\label{prop:AffineScore}
\end{prop}
This result shows that the dynamics of the transformed process $\{\overline{\bm{f}}_t\}_{t\in\mathbb{N}}$ cannot be written as a standard score-driven recursion of the form given in Equation \eqref{eq:dfm:sd:trans}.\ This is because the two matrices $\left(\bm{\mathcal{I}}_{t|t-1}\right)^{-\beta}$ and  $(\bm{T}^{-1+\beta})'$ do not generally commute, preventing the term $(\bm{T}^{-1+\beta})'$ in Equation \eqref{eq:dfm:sd:transRep} from being absorbed into the matrix $\overline{\bm{A}}$ that multiplies the scaled score. Therefore, score-driven factor models behave differently compared to parameter-driven factor models, where, for any choice of $\bm{T}$, the transformed factors  follow the same linear process describing the original factors.\ 

Among all the possible values of $\beta\in [0,1]$, there exist only two specific choices for which the term  $(\bm{T}^{-1+\beta})'$ can be absorbed into the matrix $\overline{\bm{A}}$ for any non-singular matrix $\bm{T}\in\mathbb{R}^{n\times n}$.\ We discuss these two cases below. 
\begin{itemize}
\item Case $\beta=0$.\ We have $\bm{T}^{\beta}=\bm{I}_{r\times r}$ and $\left(\overline{\bm{\mathcal{I}}}_{t|t-1}\right)^{-\beta}=\bm{I}_{r\times r}$.  Therefore, $\overline{\bm{A}}=\bm{T}^{-1}\bm{A}\bm{T}^{\beta}$ can be re-defined as $\overline{\overline{\bm{A}}}=\overline{\bm{A}}(\bm{T}^{-1})'= \bm{T}^{-1}\bm{A}(\bm{T}^{-1})'$, and Equation \eqref{eq:dfm:sd:transRep} becomes
\begin{equation}
\overline{\bm{f}}_{t+1} = \overline{\bm{c}} + \overline{\overline{\bm{A}}}\ \overline{\bm{\nabla}}_t + \overline{\bm{B}}\ \overline{\bm{f}}_t,
\end{equation}
which has the same form of Equation \eqref{eq:dfm:sd:trans} for $\beta=0$.
\item Case $\beta=1$. We have $\overline{\bm{s}}_t=\left(\overline{\bm{\mathcal{I}}}_{t|t-1}\right)^{-1}\overline{\bm{\nabla}}_t$, and therefore Equation \eqref{eq:dfm:sd:transRep} becomes
\begin{equation}
\overline{\bm{f}}_{t+1} = \overline{\bm{c}} +\overline{\bm{A}}\ \left(\overline{\bm{\mathcal{I}}}_{t|t-1}\right)^{-1}\overline{\bm{\nabla}}_t + \overline{\bm{B}}\ \overline{\bm{f}}_t,
\end{equation}
which has the same form of Equation \eqref{eq:dfm:sd:trans} for $\beta=1$.
\end{itemize}
In these two special cases, score-driven factor models are subject to the same identification issues of parameter-driven factor models because multiplying the factors by an arbitrary non-singular matrix $\bm{T}$ results in a recursion of similar form.\ For example, \cite{artemova} considers the case $\beta=1$, and uses identifying restrictions similar to those applied in parameter-driven models to infer the static parameters. 

\section{Scalar identification}
\label{sec:scalIdent}

Let us now consider the case $\beta\in (0,1)$.\ For example, we may set $\beta=\frac{1}{2}$, meaning that the score is normalized by the inverse square-root of the conditional Fisher information.\ This normalization is considered in several score-driven specifications; see \cite{creal2014observation} among others.\ Equation \eqref{eq:dfm:sd:transRep} becomes
\begin{equation}
\overline{\bm{f}}_{t+1} = \overline{\bm{c}} +\overline{\bm{A}}\ \left(\overline{\bm{\mathcal{I}}}_{t|t-1}\right)^{-\frac{1}{2}}(\bm{T}^{-\frac{1}{2}})'\overline{\bm{\nabla}}_t  + \overline{\bm{B}}\ \overline{\bm{f}}_t,
\end{equation}
which has a different structure compared to the original factor dynamics in Equation \eqref{eq:dfm:sd:trans}.\ This result can be used in order to identify the static parameters under much milder restrictions compared to those required in parameter-driven models.\ In particular, such restrictions must ensure that the two matrices $\left(\overline{\bm{\mathcal{I}}}_{t|t-1}\right)^{-\frac{1}{2}}$ and $(\bm{T}^{-\frac{1}{2}})'$ do not commute, so that  $(\bm{T}^{-\frac{1}{2}})'$ cannot be absorbed in the matrix $\overline{\bm{A}}$.\ These restrictions are summarized in the following assumptions:

\begin{ass}
The matrix $\bm{A}$ is diagonal, with non-zero diagonal elements.
\label{ass:A}
\end{ass}

\begin{ass}
The conditional Fisher information matrix $\bm{\mathcal{I}}_{t|t-1}$ is not block diagonal.
\label{ass:I}
\end{ass}

\noindent The restriction in Assumption \eqref{ass:A} is very common in the score-driven literature.\ 
Assumption \eqref{ass:I} is generally verified when the factor loading matrix $\bm{\Lambda}$ and/or the covariance matrix $\bm{\Sigma}$ are unconstrained.\ To see this, observe that, if $\bm{\epsilon}_t$ has an elliptical distribution, e.g., it is a Student-$t$, the Fisher information matrix is proportional to $\bm{\Lambda}'\bm{\Sigma}^{-1}\bm{\Lambda}$; see Appendix \ref{app:scoresInfo}.\ Therefore, $\bm{\mathcal{I}}_{t|t-1}$ is generally not block diagonal when no prior restrictions on $\bm{\Lambda}$ and/or $\bm{\Sigma}$ are imposed.\ For example, $\bm{\mathcal{I}}_{t|t-1}$ is not block diagonal when $\bm{\Lambda}$ is unrestricted  and $\bm{\Sigma}$ is diagonal. 

\begin{thm}
Let $\beta\in (0,1)$.\ Under Assumptions \eqref{ass:A} and \eqref{ass:I}, $\left(\overline{\bm{\mathcal{I}}}_{t|t-1}\right)^{-\beta}$ and $(\bm{T}^{-1+\beta})'$ commute if only if $\bm{T}$ is scalar, i.e., $\bm{T}=q\bm{I}_{r\times r}$, $q\in\mathbb{R}$. 
\label{thm:commut}
\end{thm}

\noindent The above result shows that, when the score is scaled through, e.g., the inverse square-root of the conditional Fisher information, the static parameters and the common factors are identifiable up to a common multiplicative scalar term under fairly general assumptions on the structure of the parameter space.\ Compared to parameter-driven models, where restricting the covariance of the factors to be the identity matrix leaves them defined up to a rotation, here restricting the matrix $\bm{A}$ to be diagonal identifies the factors and the static parameters up to a common multiplicative term, thus avoiding rotational indeterminacy.\ Clearly, this improves the interpretability of the model estimates because the effect of the common multiplicative factor is just to scale the factor dynamics.

To remove the indeterminacy associated with the scalar factor, it is enough to fix one parameter.\ For example, we may fix one of the entries of either $\bm{\Lambda}$ or $\bm{c}$.\ Specifically, we set
\begin{equation}
\bm{c}_1 = 1,
\end{equation} 
which implies $q=1$.\ This choice is more convenient than setting a constraint for the factor loadings because the latter would break the structure of $\bm{\Lambda}$ when the order of the time-series changes.\ On the contrary, setting $\bm{c}_1 = 1$ leads to order-invariant restrictions.\ This result holds for the wide class of elliptical distributions, as shown in the next proposition.

\begin{prop}
Let us assume that $p(\bm{y}_t| \bm{\mathcal{F}}_{t-1},\bm{f}_t)$ is in the class of elliptical distributions, that is, $p(\bm{y}_t|\bm{\mathcal{F}}_{t-1},\bm{f}_t)=(\det{\bm{\Sigma}})^{-\frac{1}{2}}\psi ((\bm{y}_t-\bm{\Lambda}\bm{f}_t)'\bm{\Sigma}^{-1}(\bm{y}_t-\bm{\Lambda}\bm{f}_t))$, where $\psi$ is a scalar function.\ We further assume that $\bm{\Lambda}$ is unconstrained.\ Then the likelihood is invariant for transformations of the form $\bm{y}_t\to \bm{P}\bm{y}_t$, where $\bm{P}$ is a permutation matrix.\ Furthermore, the filtered dynamics of $\bm{f}_t$ are not affected by the permutation if the same starting value is used to initialize the filter of the permuted series.
\label{prop:order}
\end{prop}

\section{Generalizations}
\label{sec:gener}

\subsection{Score-driven factor loadings}
\label{sec:tvLoad}
One of the advantages of the normalization based on the inverse square-root of the conditional Fisher information is that, under the same restrictions specified in Assumptions \eqref{ass:A}, \eqref{ass:I}, it is possible to resolve the rotational indeterminacy also in factor models with time-varying loadings.\ Let us consider the following dynamic factor model
\begin{equation}
\bm{y}_t = \bm{\Lambda}_t \bm{g}_t + \bm{\epsilon}_t
\label{eq:dfm:sd:obs_tv}
\end{equation}
where $\bm{\Lambda}_t$, $\bm{g}_t$ are both predictable with respect to $\bm{\mathcal{F}}_t$.\  Let $\bm{l}_t=\text{vec}(\bm{\Lambda}_t)$ be an $n r\times 1 $ vector obtained by stacking the elements of $\bm{\Lambda}_t$ in a column.\ The score-driven dynamics of $\bm{l}_t$ and $\bm{g}_t$  are given by
\begin{align}
\bm{l}_{t+1}&=\bm{c}^{(l)}+ \bm{A}^{(l)} \bm{s}_t^{(l)} + \bm{B}^{(l)}\bm{l}_t\label{eq:gas_l}\\
\bm{g}_{t+1}&=\bm{c}^{(g)}+ \bm{A}^{(g)} \bm{s}_t^{(g)} + \bm{B}^{(g)}\bm{g}_t\label{eq:gas_g}
\end{align}
where 
\begin{align}
\bm{s}_t^{(l)} &= \bm{S}_t^{(l)} \bm{\nabla}_t^{(l)},\quad\quad \bm{\nabla}_t^{(l)} = \left[\frac{\partial\log p(\bm{y}_t|\bm{\mathcal{F}}_{t-1},\bm{l}_t,\bm{g}_t)}{\partial\bm{l}_t}\right]'\\
\bm{s}_t^{(g)} &= \bm{S}_t^{(g)} \bm{\nabla}_t^{(g)},\quad\quad \bm{\nabla}_t^{(g)} =\left[ \frac{\partial\log p(\bm{y}_t|\bm{\mathcal{F}}_{t-1},\bm{l}_t,\bm{g}_t)}{\partial\bm{g}_t}\right]'.
\end{align}
The two normalization matrices $\bm{S}_t^{(l)}$, $\bm{S}_t^{(g)}$ are set as follows
\begin{equation}
\bm{S}_t^{(l)} =\left(\bm{\mathcal{I}}_{t|t-1}^{(l)}\right)^{-\alpha},\quad\quad \bm{S}_t^{(g)} =\left(\bm{\mathcal{I}}_{t|t-1}^{(g)}\right)^{-\beta}\medskip
\end{equation} 
where $\alpha,\beta\in\ [0,1]$, and $\bm{\mathcal{I}}_{t|t-1}^{(l)}$ and $\bm{\mathcal{I}}_{t|t-1}^{(g)}$ are two conditional Fisher Information matrices
\begin{equation}
\bm{\mathcal{I}}_{t|t-1}^{(l)} = \mathbb{E}[\bm{\nabla}_t^{(l)}\bm{\nabla}_t^{(l)\prime}|\bm{\mathcal{F}}_{t-1}],\quad\quad \bm{\mathcal{I}}_{t|t-1}^{(g)} = \mathbb{E}[\bm{\nabla}_t^{(g)}\bm{\nabla}_t^{(g)\prime}|\bm{\mathcal{F}}_{t-1}]
\end{equation}

Proposition \eqref{prop:AffineScore} can be generalized as follows:
\begin{prop}
Let $\bm{T}\in\mathbb{R}^{r\times r}$ be non-singular and let us set $\overline{\bm{\Lambda}}_t=\bm{\Lambda}_t\bm{T}$, $\overline{\bm{l}}_t=\emph{vec}(\bm{\overline{\Lambda}}_t)=(\bm{T}'\otimes \bm{I})\emph{vec}(\bm{\Lambda}_t)$, $\overline{\bm{g}}_t=\bm{T}^{-1}\bm{g}_t$. The score-driven factor model with time-varying loadings in Equations \eqref{eq:dfm:sd:obs_tv}, \eqref{eq:gas_l}, \eqref{eq:gas_g} can be re-parameterized as follows:
\begin{align}
\bm{y}_t &= \overline{\bm{\Lambda}}_t\ \overline{\bm{g}}_t + \bm{\epsilon}_t\label{eq:dfm:sd:obsRep_tv}\\
\overline{\bm{l}}_{t+1} &= \overline{\bm{c}}^{(l)}+ \overline{\bm{A}}^{(l)}  \left(\overline{\bm{\mathcal{I}}}_{t|t-1}^{(l)}\right)^{-\alpha}(\bm{T}^{-\alpha+1}\otimes \bm{I})\overline{\bm{\nabla}}_t^{(l)}       + \overline{\bm{B}}^{(l)} \overline{\bm{l}}_t
\label{eq:gas_l_transf_tv}\\
\overline{\bm{g}}_{t+1} &=\overline{\bm{c}}^{(g)}+\overline{\bm{A}}^{(g)}  \left(\overline{\bm{\mathcal{I}}}_{t|t-1}^{(g)}\right)^{-\beta}(\bm{T}^{-1+\beta})^{\prime}\overline{\bm{\nabla}}_t^{(g)}   +\overline{\bm{B}}^{(g)}\overline{\bm{g}}_t\medskip
\label{eq:gas_g_transf_tv}
\end{align}
where $\overline{\bm{c}}^{(l)}=(\bm{T}'\otimes \bm{I})\bm{c}^{(l)}$,  $\overline{\bm{A}}^{(l)}=(\bm{T}'\otimes \bm{I})\bm{A}^{(l)}(\bm{T}'\otimes \bm{I})^{-\alpha}$, $\overline{\bm{B}}^{(l)}=(\bm{T}'\otimes \bm{I})\bm{B}^{(l)}(\bm{T}'\otimes \bm{I})^{-1}$,
$\overline{\bm{c}}^{(g)}=\bm{T}^{-1}\bm{c}^{(g)}$, $\overline{\bm{A}}^{(g)}=\bm{T}^{-1}\bm{A}^{(g)}\bm{T}^{\beta}$, $\overline{\bm{B}}^{(g)}=\bm{T}^{-1}\bm{B}^{(g)}\bm{T}$, and\medskip
\begin{align}
\overline{\bm{\nabla}}_t^{(l)}&=\left[\frac{\partial\log\mathcal{L}(\bm{y}_t|\bm{\mathcal{F}}_{t-1},\overline{\bm{l}}_t,\overline{\bm{g}}_t)}{\partial\bm{\overline{l}}_t}\right]',  \quad \overline{\bm{\mathcal{I}}}_{t|t-1}^{(l)}= \mathbb{E}[\overline{\bm{\nabla}}_t^{(l)}\overline{\bm{\nabla}}_t^{(l)\prime}|\bm{\mathcal{F}}_{t-1}]\\
\overline{\bm{\nabla}}_t^{(g)}&=\left[\frac{\partial\log\mathcal{L}(\bm{y}_t|\bm{\mathcal{F}}_{t-1},\overline{\bm{l}}_t,\overline{\bm{g}}_t)}{\partial\bm{\overline{g}}_t}\right]',  \quad \overline{\bm{\mathcal{I}}}_{t|t-1}^{(g)}= \mathbb{E}[\overline{\bm{\nabla}}_t^{(g)}\overline{\bm{\nabla}}_t^{(g)\prime}|\bm{\mathcal{F}}_{t-1}]
\end{align}
\label{prop:AffineScore_tv}
\end{prop}
Also in this case, the dynamics of the transformed process $\overline{\bm{g}}_t$ cannot be represented as a score-driven recursion.\
 In particular, for $\beta\in (0,1)$, it is still true that the law of motion of $\overline{\bm{g}}_t$ has a different structure relative to Equation \eqref{eq:gas_g} due to the non-commutativity of the two matrices $(\bm{T}^{-1+\beta})'$ and $\left(\overline{\bm{\mathcal{I}}}_{t|t-1}^{(g)}\right)^{-\beta}$.\ Compared to the static case presented in the previous section, the main difference here is that the Fisher information $\overline{\bm{\mathcal{I}}}_{t|t-1}^{(g)}$ is generally time-varying.\ For instance, when the distribution of $\bm{\epsilon}_t$ is in the class of elliptical distributions, $\overline{\bm{\mathcal{I}}}_{t|t-1}^{(g)}$ is proportional to $\overline{\bm{\Lambda}}_t'\bm{\Sigma}^{-1}\overline{\bm{\Lambda}}_t$.\ The following two assumptions, which are the analogue of Assumptions \eqref{ass:A}, \eqref{ass:I}, are sufficient, also in this more general case, to prevent the two matrices $(\bm{T}^{-1+\beta})'$ and $\left(\overline{\bm{\mathcal{I}}}_{t|t-1}^{(g)}\right)^{-\beta}$ to commute.

\begin{ass}
The matrix $\bm{A}^{(g)}$ is diagonal, with non-zero diagonal elements.
\label{ass:A_tv}
\end{ass}

\begin{ass}
The conditional Fisher information matrix $\bm{\mathcal{I}}_{t|t-1}^{(g)}$ is not block diagonal.
\label{ass:I_tv}
\end{ass}

\begin{thm}
Let $\beta\in (0,1)$.\ Under Assumptions \eqref{ass:A_tv} and \eqref{ass:I_tv}, $\left(\overline{\bm{\mathcal{I}}}_{t|t-1}\right)^{-\beta}$ and $(\bm{T}^{-1+\beta})'$ commute if only if $\bm{T}$ is scalar, i.e., $\bm{T}=q\bm{I}_{r\times r}$, $q\in\mathbb{R}$. 
\label{thm:commut_tv}
\end{thm}

The main consequence of Theorem \eqref{thm:commut_tv} is that we can identify the model parameters up to a common multiplicative scalar term also in a dynamic setting with time-varying loadings.\ This is true because Assumptions \eqref{ass:A_tv} and \eqref{ass:I_tv} restrict the matrix $\bm{T}$ to be scalar, and thus also the parameters governing the dynamics of $\bm{\Lambda}_t$ can be identified up to a common scalar term. 

We use the same identifying restriction presented in the static case, i.e., we fix the first element of $\bm{c}^{(l)}$
\begin{equation}
\bm{c}_1^{(l)} = 1,
\end{equation} 
which implies $q=1$.\ It is worth to note that $q$ is uniquely determined independently from the normalization $\alpha\in [0,1]$ adopted to scale the score of the factor loading dynamics.\ In our simulation and empirical analysis, we set $\alpha=0$ since $\overline{\bm{\mathcal{I}}}_{t|t-1}^{(l)}$ is generally a singular matrix in models characterized by an elliptical distribution for the idiosyncratic noise $\bm{\epsilon}_t$.

\subsection{Nonlinear factor models}
Let $h:\mathbb{R}^r\to\mathbb{R}^d$ be a differentiable vector function. We consider the following nonlinear factor specification
\begin{align}
\bm{y}_t |\bm{\mathcal{F}}_{t-1} &\sim p(\bm{y}_t|\bm{\mathcal{F}}_{t-1},\bm{h}(\bm{f}_t)) \label{eq:dfm:sd:obs:nl}\\
\bm{h}(\bm{f}_t)&=\bm{\Lambda}\bm{f}_t\\
\bm{f}_{t+1} &= \bm{c} +\bm{A}\bm{s}_t + \bm{B}\bm{f}_t\label{eq:dfm:sd:trans:nl}
\end{align}
where $p(\bm{y}_t|\bm{\mathcal{F}}_{t-1},\bm{h}(\bm{f}_t))$ denotes the conditional probability density of $\bm{y}_t$ given a time-varying vector $\bm{h}(\bm{f}_t)\in\mathbb{R}^d$, $\bm{\Lambda}\in\mathbb{R}^{d\times r}$, and $\bm{f_t}\in\mathbb{R}^r$ is a vector of common factors driving the dynamics of the time-varying parameters.\ Here, $\bm{s}_t$ is still the normalized score, i.e., $\bm{s}_t=\bm{S}_t\bm{\nabla}_t$, $\bm{S}_t=(\bm{\mathcal{I}}_{t|t-1})^{-\beta}$, where 
\begin{equation}
\bm{\nabla}_t = \left[\frac{\partial\log p(\bm{y}_t|\bm{\mathcal{F}}_{t-1},\bm{h}(\bm{f}_t))}{\partial\bm{f}_t}\right]',\qquad \bm{\mathcal{I}}_{t|t-1}=\mathbb{E}[\bm{\nabla}_t\bm{\nabla}_t'|\bm{\mathcal{F}}_{t-1}].
\end{equation}
It is straightforward to verify that multiplying $\bm{f}_t$ by a non-singular matrix results in the same transformation described in Proposition \eqref{prop:AffineScore}.\ Therefore, we can adopt the restrictions described in Section \eqref{sec:scalIdent} to identify the static parameters.\ In general, the factor structure in models with nonlinear observation density has the role of reducing the number of parameters, which might otherwise increase very fast with the dimensionality.\
Some examples of nonlinear score-driven factor models are given in \cite{creal2011dynamic}, who propose common factors driving the time-varying correlations in order to reduce the model dimensionality,  and \cite{opschoor2021closed}, who consider a copula model with dynamic common factors.\ 

\section{Monte Carlo analysis}
\label{sec:MC}

In this section, we perform an extensive Monte Carlo analysis to demonstrate that the model is identifiable under the scalar restriction discussed in Section \eqref{sec:scalIdent}.\ To this end, we study the finite sample properties of the standard maximum-likelihood estimator in both the constant and time-varying loading cases.
Throughout the study, we use the following Student-$t$ conditional density to derive the score-driven law of motion of the time-varying parameters:
\begin{equation}
p(\bm{y}_t|\bm{\mathcal{F}}_{t-1},\bm{f}_t) = 
\frac{\Gamma\left(\frac{\nu+n}{2}\right)}{\Gamma\left(\frac{\nu}{2}\right)((\nu-2)\pi)^{\frac{n}{2}}\text{det}(\bm{\Sigma})^{1/2}}\left[1+  \frac{1}{\nu-2}(\bm{y}_t-\bm{\Lambda}\bm{g}_t)'\bm{\Sigma}^{-1}(\bm{y}_t-\bm{\Lambda}\bm{g}_t) \right]^{-\frac{\nu+n}{2}},
\label{eq:student-t}
\end{equation}
where $\bm{\Lambda}$ is replaced by $\bm{\Lambda}_t$ in the time-varying loading case.\ Explicit expressions of the scores and conditional Fisher information for both $\bm{f}_t$ and $\bm{l}_t=\text{vec}(\bm{\Lambda}_t)$ are given in Appendix \ref{app:scoresInfo}.

\subsection{Static loadings}
\label{sub:MC1}
We first consider a low-dimensional setting with $n=5$ time series and $r=2$ common factors.\ The parameters governing the score-driven factor dynamics are set as follows:
\begin{equation}
\bm{c}=
\begin{pmatrix}
1\\
0.1
\end{pmatrix},\quad\quad
\bm{A}=
\begin{pmatrix}
0.1 & 0\\
0 & 0.3
\end{pmatrix},\quad\quad
\bm{B}=
\begin{pmatrix}
0.9 & 0\\
0 & 0.7
\end{pmatrix}.
\end{equation}
Moreover, we set $\nu=5$, $\bm{\Sigma}=\frac{1}{2}\bm{I}_{5\times 5}$, $\bm{\Lambda}_{ij}\sim\text{Uniform}(0,1)$, for $i=1,\dots,5$ and $j=1,2$.\ The first entry of $\bm{c}$ is kept fixed in order to resolve the scalar invariance and identify all the parameters.\ The covariance matrix $\bm{\Sigma}$ is scalar in this example, however no scalar restriction is imposed in the estimation, i.e., we estimate each diagonal entry of $\bm{\Sigma}$ independently from others.\ Note that, with this choice of the static parameters, our restrictions in Assumptions \ref{ass:A}, \ref{ass:I} are satisfied.\ The initial values of the static parameters given as an input to the likelihood optimization are obtained by perturbing the true values with uniform random shifts ranging from 0 to 50\%.

\begin{figure}[h!]
    \centering 
\begin{subfigure}{0.3\textwidth}
  \includegraphics[width=\linewidth]{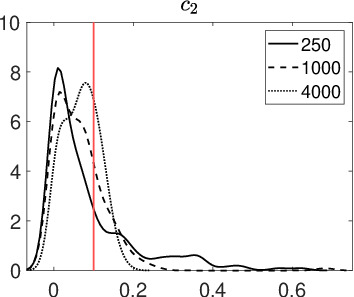}
\end{subfigure}\hfil 
\begin{subfigure}{0.3\textwidth}
  \includegraphics[width=\linewidth]{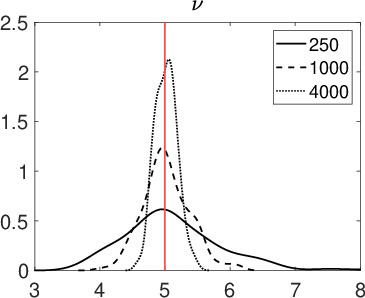}
\end{subfigure}\hfil 
\begin{subfigure}{0.3\textwidth}
  \includegraphics[width=\linewidth]{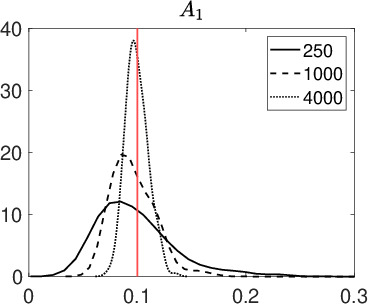}

\medskip  
  
\end{subfigure}\hfil 
\begin{subfigure}{0.3\textwidth}
  \includegraphics[width=\linewidth]{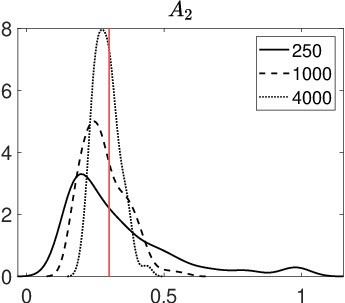}
\end{subfigure}\hfil 
\begin{subfigure}{0.3\textwidth}
  \includegraphics[width=\linewidth]{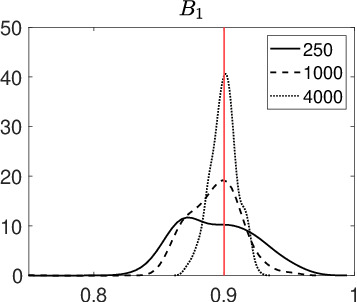}
\end{subfigure}\hfil 
\begin{subfigure}{0.3\textwidth}
  \includegraphics[width=\linewidth]{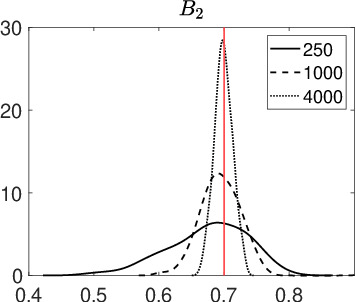}
\end{subfigure}\hfil 

\caption{Kernel density estimates of the maximum likelihood estimates of $\nu$ and each individual element of $\bm{c}$, $\bm{A}$, $\bm{B}$. The estimates are obtained based on $N=250$ replications of a score-driven factor model with sample size $T=250,1000,4000$.\ The vertical red line denotes the true parameter value.\ The results for the remaining parameters are reported in the Appendix. }
\label{fig:MC1}
\end{figure}

We generate $N=250$ replications of the model with different sample sizes $T=250,1000,4000$.\ The kernel density estimates of the maximum-likelihood estimates of $\nu$ and of each individual element of $\bm{c}$, $\bm{A}$, $\bm{B}$ are shown in Figure \eqref{fig:MC1}, whereas those of $\bm{\Lambda}$ and $\bm{\Sigma}$ are reported in Appendix \ref{app:sub:MC1} to save space.\ The results show that the maximum-likelihood estimator concentrates around the true values as the sample size increases, thus confirming that the model parameters are identifiable under the restriction.

\begin{figure}[h!]
    \centering 
\begin{subfigure}{0.3\textwidth}
  \includegraphics[width=\linewidth]{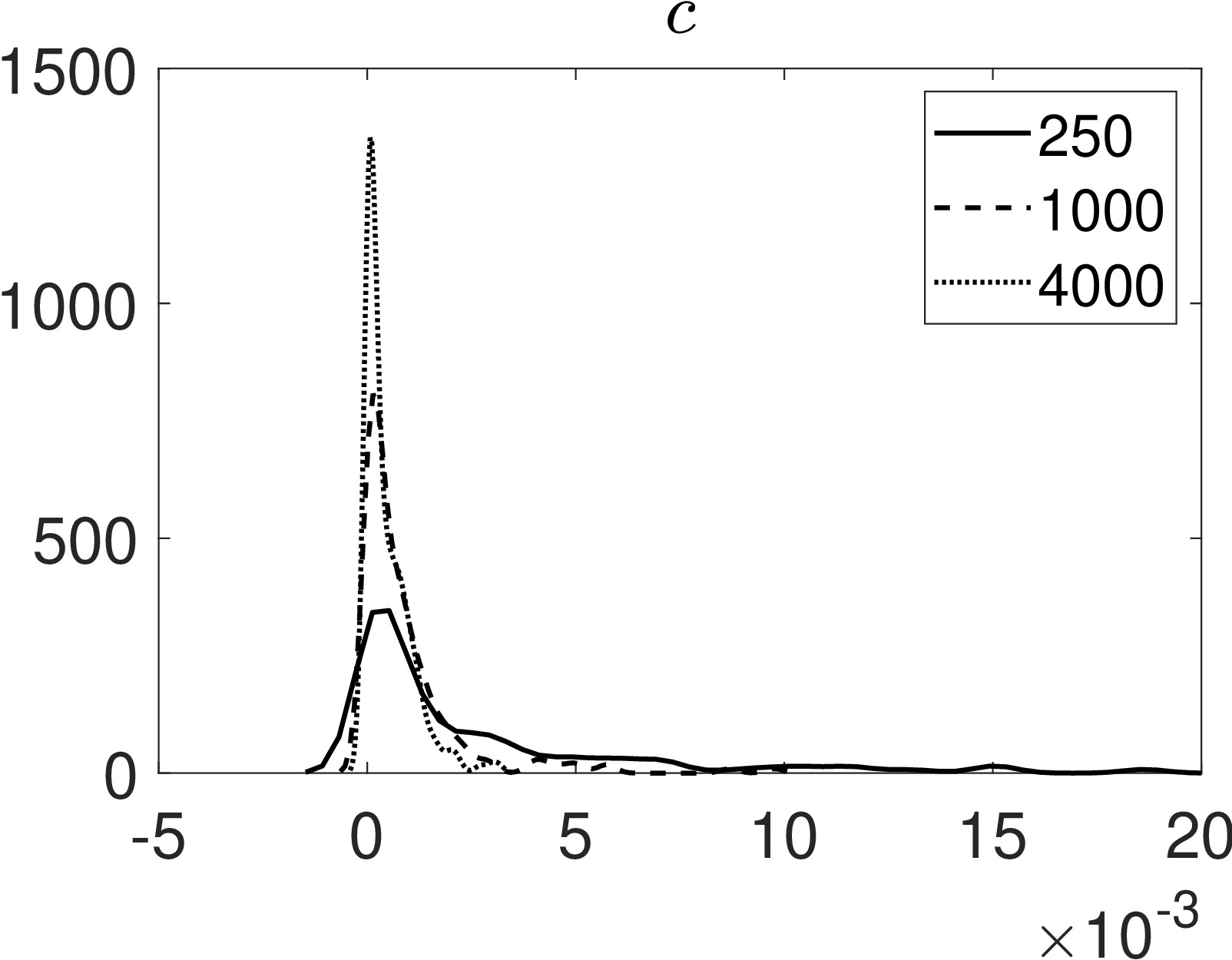}
\end{subfigure}\hfil 
\begin{subfigure}{0.3\textwidth}
  \includegraphics[width=\linewidth]{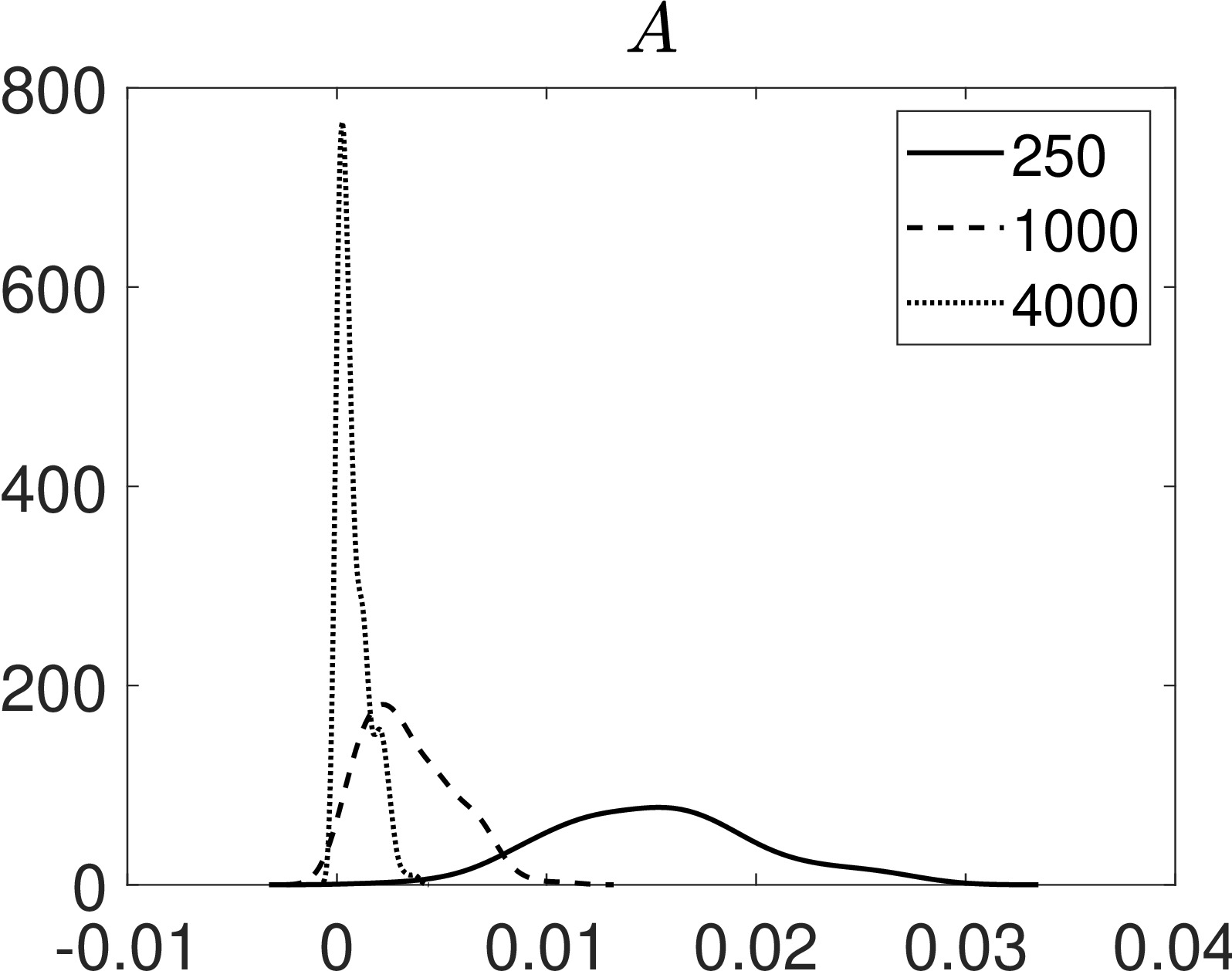}
\end{subfigure}\hfil 
\begin{subfigure}{0.3\textwidth}
  \includegraphics[width=\linewidth]{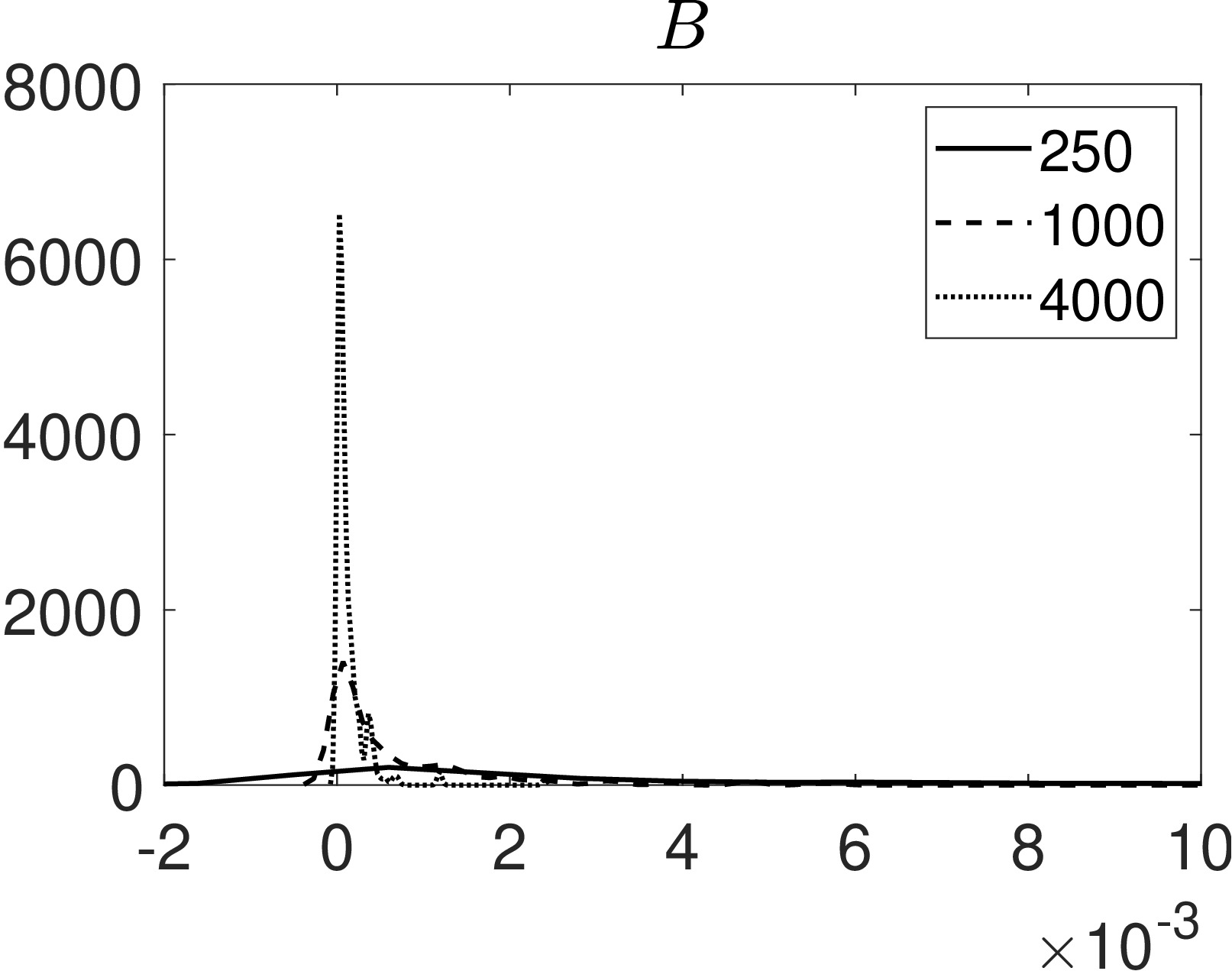}

\medskip  
  
\end{subfigure}\hfil 
\begin{subfigure}{0.3\textwidth}
  \includegraphics[width=\linewidth]{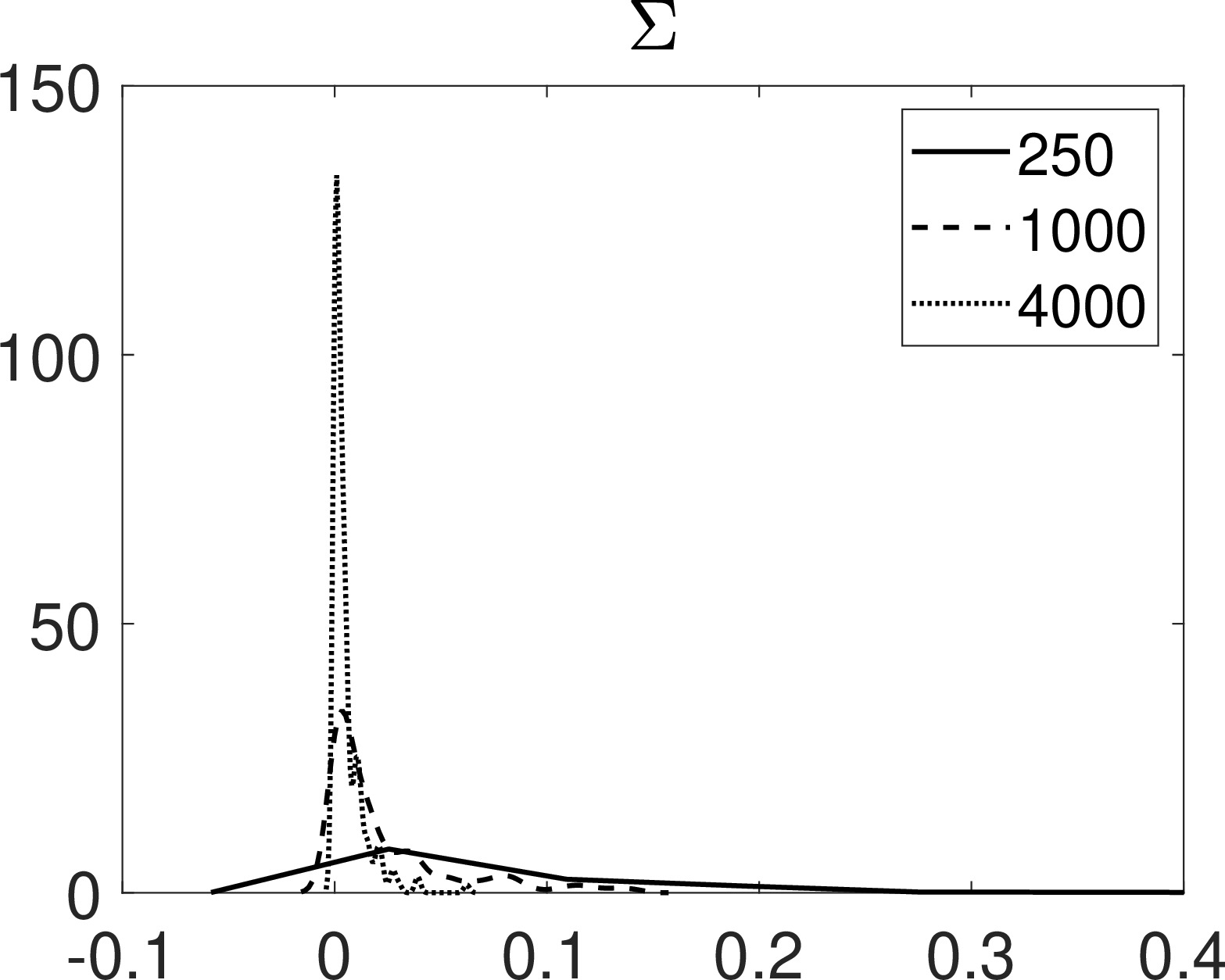}
\end{subfigure}\hfil 
\begin{subfigure}{0.3\textwidth}
  \includegraphics[width=\linewidth]{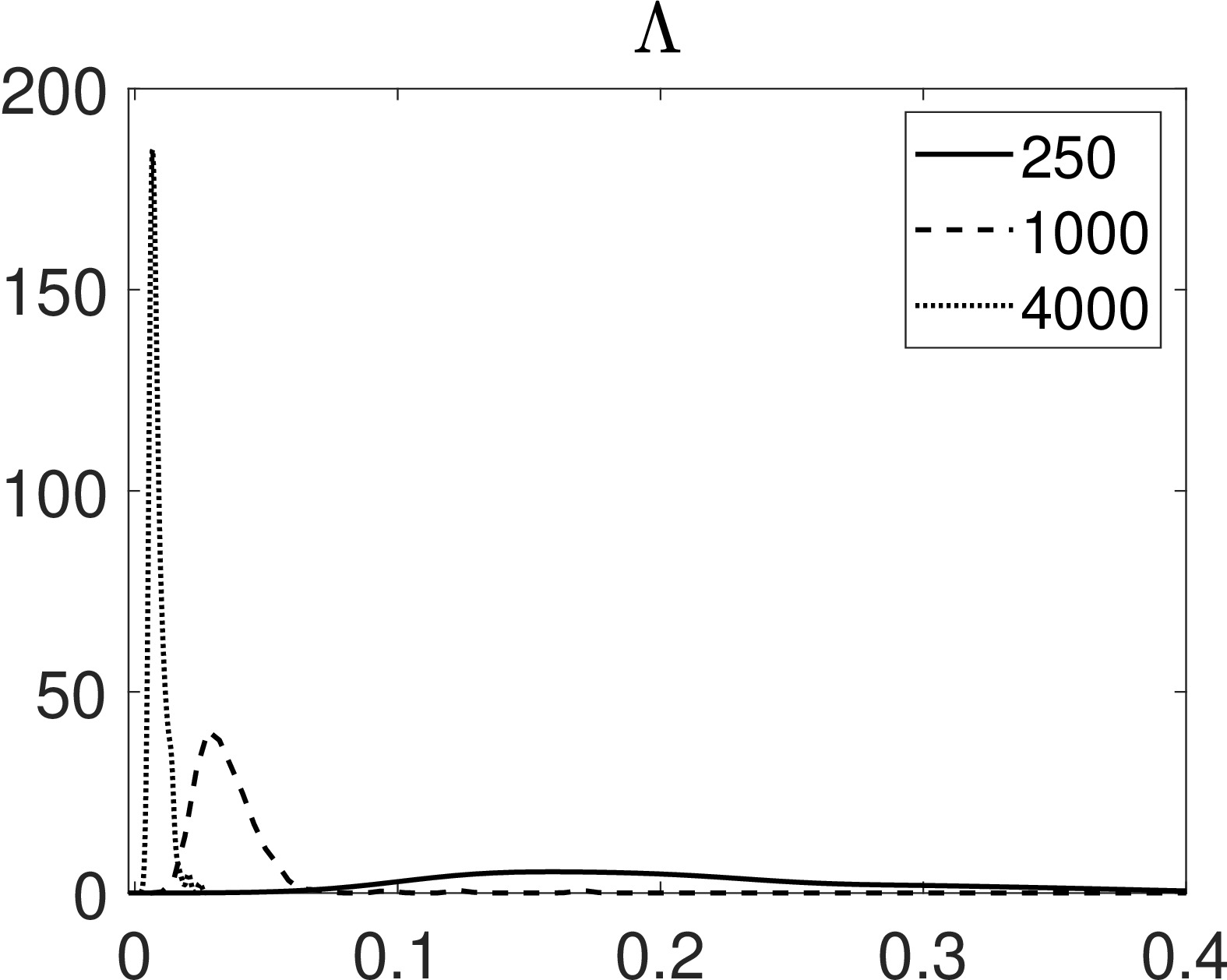}
\end{subfigure}\hfil 
\begin{subfigure}{0.3\textwidth}
  \includegraphics[width=\linewidth]{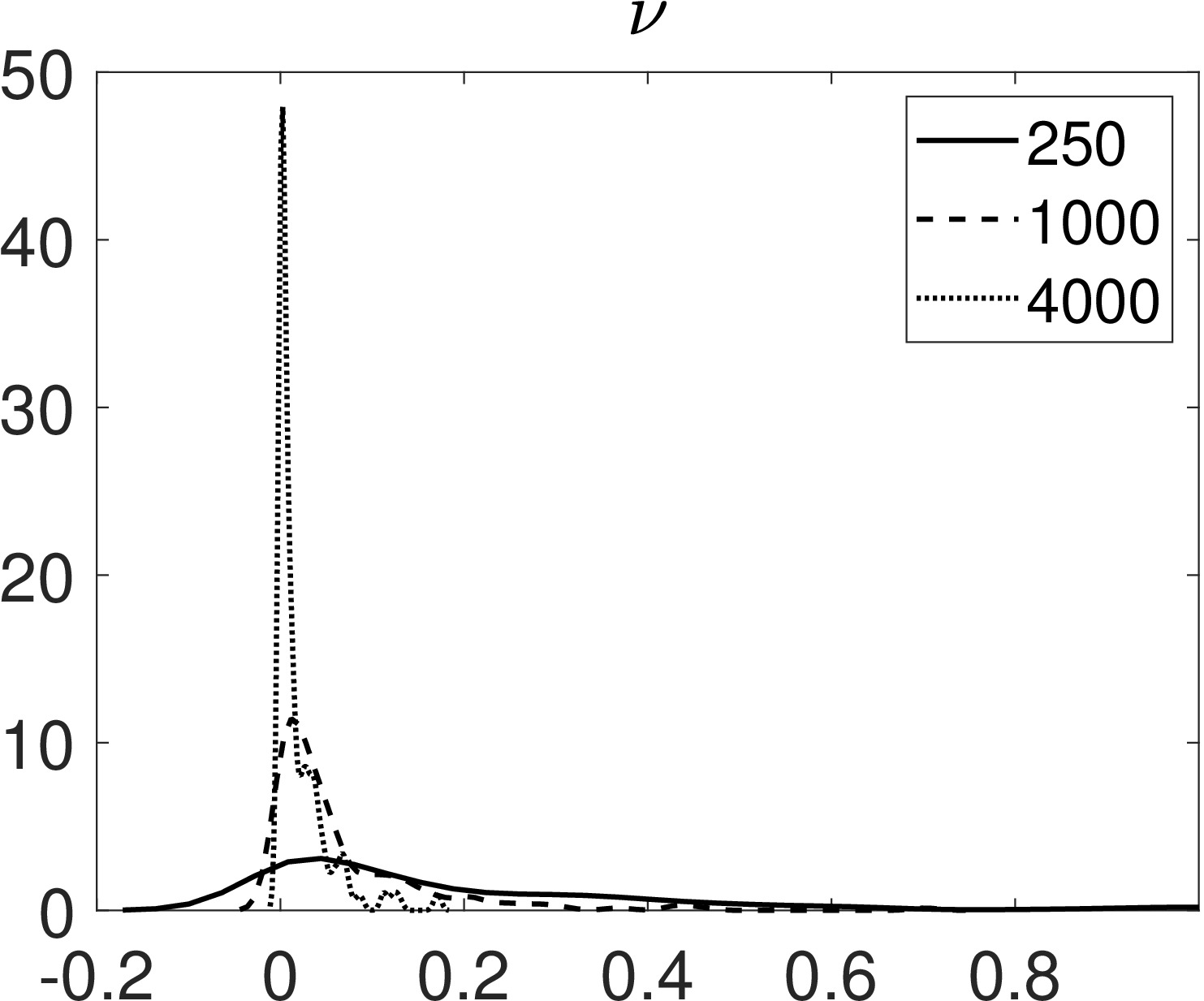}
\end{subfigure}\hfil 

\caption{Kernel density estimates of Frobenius distance between the true model matrices $\bm{c}$, $\bm{A}$, $\bm{B}$, $\bm{\Lambda}$, $\bm{\Sigma}$, $\nu$ and their maximum-likelihood  estimates for $T=250,1000,4000$.\ The estimates are obtained based on $N=250$ replications of a score-driven factor model with dimension $n=100$ and $r=2$ factors. }
\label{fig:MC3}
\end{figure}

We now consider a large dimensional setting with $n=100$ and $r=2$.\ The parameters $\bm{c}$, $\bm{A}$, $\bm{B}$, $\nu$ are chosen as in the previous study, whereas $\bm{\Sigma}$ and $\bm{\Lambda}$ are set as  $\bm{\Sigma}=2\bm{I}_{100\times 100}$, and $\bm{\Lambda}_{ij}\sim\text{Uniform}(0,1)$, for $i=1,\dots,100$ and $j=1,2$.\ Note that the variance of the idiosyncratic noise is increased in order to achieve a signal-to-noise ratio comparable to that used in the previous setting.\ We generate $N=250$ replications of the model with $T=[250,1000,4000]$. 
Figure \eqref{fig:MC3} shows the kernel density estimates of the Frobenius distances between the true matrices and the maximum-likelihood estimates.\ For all model matrices, we note a significant drop in the Frobenius norms as the sample size increases, indicating that the estimates get closer to the true parameter values as the sample size increases.\

\subsection{Time-varying loadings}
\label{sub:MC4}
As a second exercise, we study the finite-sample behavior of the score-driven factor model with time-varying coefficients in order to verify that the model parameters are  correctly estimated under the same set of identifying restrictions.\ We consider a simulation setting with $n=5$ time-series variables and $r=2$ common factors.\ The parameters $\bm{c}^{(g)}$, $\bm{A}^{(g)}$, $\bm{B}^{(g)}$, $\bm{\Sigma}$, $\nu$ are set as in Section \eqref{sub:MC1}.\ The parameters governing the loading dynamics are instead set as follows: $\bm{c}_i^{(l)}=0.1$, $\bm{A}_{ii}^{(l)}\sim \text{Uniform}(0,0.5)$, $\bm{B}_{ii}^{(l)}=0.9$, for $i=1,\dots,10$.\ In this setting, the two matrices $\bm{\Sigma}$, $\bm{\bm{B}}$ are scalar, however each diagonal element is estimated independently from others.\ In contrast, we estimate the same constant $\bm{c}_i^{(l)}$ for all the time-varying loadings.\ Other configurations are possible, for example we may estimate the same $\bm{A}_{ii}^{(l)}$ and $\bm{B}_{ii}^{(l)}$ for any loading and a different $\bm{c}_i^{(l)}$, or we may choose different $\bm{c}_i^{(l)}$, $\bm{A}_{ii}^{(l)}$ and $\bm{B}_{ii}^{(l)}$ for all loadings.\ The results we obtain in these different configurations do not different substantially from each others, and thus to save space we report here those related to the first configuration with different $\bm{A}_{ii}^{(l)}$ and $\bm{B}_{ii}^{(l)}$  and common $\bm{c}_i^{(l)}$.

The initial parameter values for the likelihood maximization are set as in Section \eqref{sub:MC1}, i.e., by perturbing the true values with uniform random shifts ranging from 0 to 50\%.\ The kernel density estimates of $\bm{c}^{(g)}$, $\bm{A}^{(g)}$, $\bm{B}^{(g)}$, $\nu$ are reported in Figure \eqref{fig:MC4}, while the results for the remaining parameters are shown in Appendix \ref{app:sub:MC4}.\ We note that, similarly to the static case in Section \eqref{sub:MC1}, all the static parameters are correctly estimated and the distribution of the maximum-likelihood estimator concentrates around the true values as the sample size increases.\ This confirms the results of Section \eqref{sec:tvLoad}, showing that the model parameters are identifiable under the scalar restriction also when the factor loading matrix is time-varying.

\begin{figure}[h!]
    \centering 
\begin{subfigure}{0.3\textwidth}
  \includegraphics[width=\linewidth]{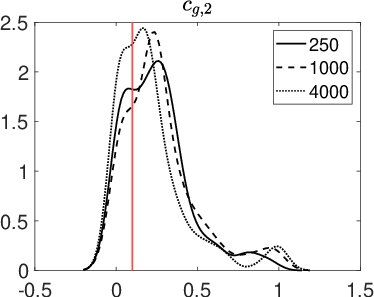}
\end{subfigure}\hfil 
\begin{subfigure}{0.3\textwidth}
  \includegraphics[width=\linewidth]{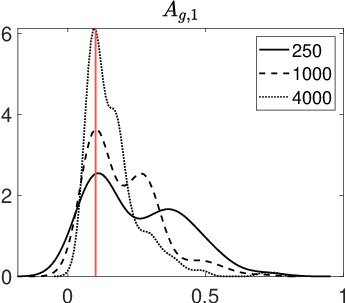}
\end{subfigure}\hfil 
\begin{subfigure}{0.3\textwidth}
  \includegraphics[width=\linewidth]{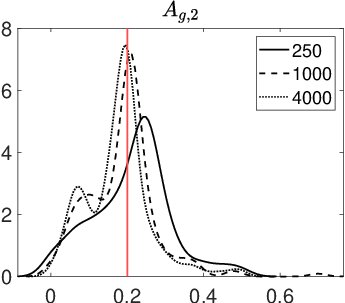}

\medskip  
  
\end{subfigure}\hfil 
\begin{subfigure}{0.3\textwidth}
  \includegraphics[width=\linewidth]{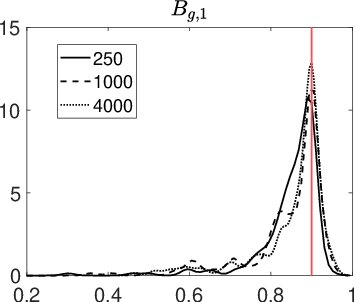}
\end{subfigure}\hfil 
\begin{subfigure}{0.3\textwidth}
  \includegraphics[width=\linewidth]{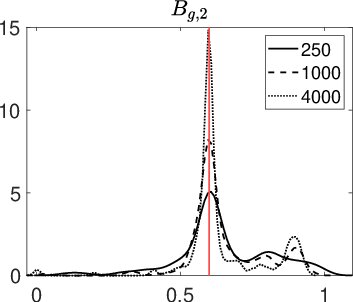}
\end{subfigure}\hfil 
\begin{subfigure}{0.3\textwidth}
  \includegraphics[width=\linewidth]{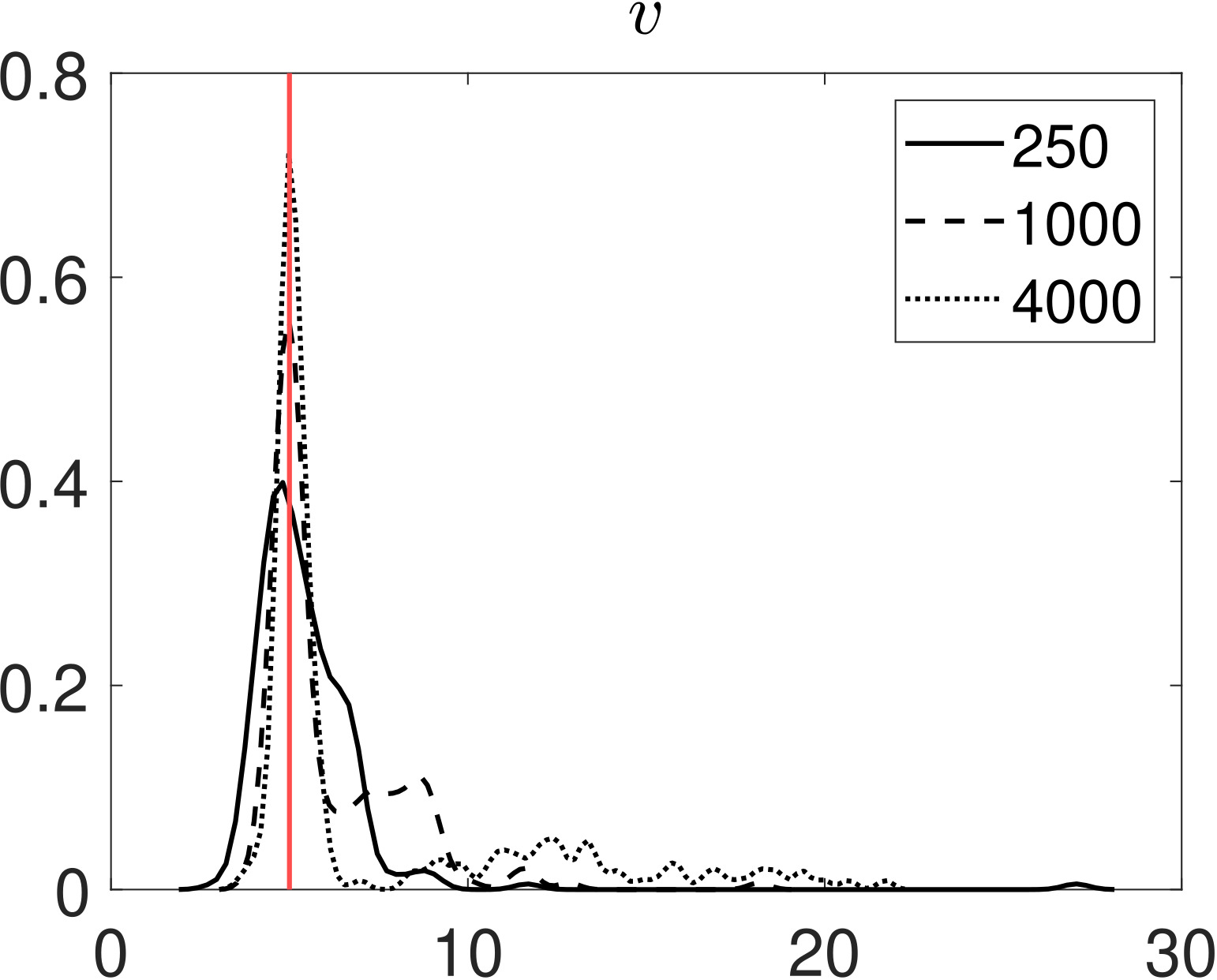}
\end{subfigure}\hfil 

\caption{Kernel density estimates of the maximum likelihood estimates of $\nu$ and each individual element of $\bm{c}^{(g)}$, $\bm{A}^{(g)}$, $\bm{B}^{(g)}$. The estimates are obtained based on $N=250$ replications of a score-driven factor model with time-varying loadings with sample size $T=250,1000,4000$.\ The vertical red line denotes the true parameter value.\ The results for the remaining parameters are reported in the Appendix. }
\label{fig:MC4}
\end{figure}

\section{Empirical Application}
\label{sec:EA}

In the following empirical analyses, the main goal is to assess the potential advantages of the flexible dependence structure achieved through the introduced identification scheme to capture the dynamics of the financial and macroeconomic time series. To this end, we apply the proposed framework to two datasets: (i) a panel of 8 macro-financial time series previously  applied in the literature to extract economic activity indicators (e.g., \citealt{creal2014observation}, \citealt{artemova}) and (ii) the daily returns series of 87 constituents of the S\&P500.

In each application, we compare the benchmark score-driven factor model given by Equations \eqref{eq:dfm:sd:obs} and \eqref{eq:dfm:sd:trans} under the scale \(\beta=\frac{1}{2}\), fixed constant \(c_1\), and unrestricted loading matrix \(\bm{\Lambda}\), with a set of specifications with  restricted loading matrix. In particular, we initially consider the common restrictions from the earlier literature to insure the identifiability of the model parameters; see, e.g., \cite{bai2012statistical} and \cite{artemova2022score}. In this regard, the lower-tringular structure is imposed on the first \(r\) rows of the loading matrix corresponding to the \(r\) factors, with the diagonal elements set equal to 1. 
To illustrate, let us consider \(N=12\) assets exposed to \(r=3\) latent factors. The resulting \(N \times r\) matrix of loadings, labeled as general lower-triangular (LT), is given by:

\begin{equation}
\bm{\Lambda}=
\begin{pmatrix}
1 & 0 & 0\\
\lambda_{2,1} & 1 & 0\\
\lambda_{3,1} & \lambda_{3,2} & 1\\
\lambda_{4,1} & \lambda_{4,2} & \lambda_{4,3}\\
\vdots & \vdots & \vdots\\
\lambda_{12,1} & \lambda_{12,2} & \lambda_{12,3}\\
\end{pmatrix}.
\label{eq:lt}
\end{equation}

Then, we examine the loading structures in score-driven factor copula models of \cite{opschoor2021closed}, ensuring the identification of the model parameters. They divide the assets into \(p\) groups according to their industry classification, assuming that all the assets within groups share the analogous dependence structure. As such, let us assume that \(N=12\) assets belong to \(p=3\) equally-sized groups. The first loading matrix is restricted to the lower-triangular form of group-specific loadings to \(r=p\) factors

\begin{equation}
\bm{\Tilde{\Lambda}}=
\begin{pmatrix}
\Tilde{\lambda}_{1,1} & 0 & 0\\
\Tilde{\lambda}_{2,1} & \Tilde{\lambda}_{2,2} & 0\\
\Tilde{\lambda}_{3,1} & \Tilde{\lambda}_{3,2} &\Tilde{\lambda}_{3,3}\\
\end{pmatrix},
\label{eq:gs-lt}
\end{equation}
where each \(\Tilde{\lambda}_{i,j}\), for the group \(i=1,2,3,\) and factor \(j=1,2,3,\) is a \(4 \times 1\) vector. 

The second restriction implies the common exposure of all groups to the so-called common factor and individual loadings on the group-specific one for a total of \(p+1\) unique factor loadings. In the example above  with \(p=3\) groups and \(r=p+1=4\) factors, the restrictions would be:
\begin{equation}
\bm{\Tilde{\Lambda}}=
\begin{pmatrix}
\Tilde{\lambda}_{1} & \Tilde{\lambda}_{1,2} & 0 & 0\\
\Tilde{\lambda}_{1} & 0 & \Tilde{\lambda}_{2,3} & 0\\
\Tilde{\lambda}_{1} & 0 & 0 & \Tilde{\lambda}_{3,4}\\
\end{pmatrix}.
\label{eq:gs}
\end{equation}

Ultimately, the third specification features all the groups exposed to \(r=2\) factors with the common and group-specific loadings on the first and the second factor, respectively. The resulting loading matrix is given by:

\begin{equation}
\bm{\Tilde{\Lambda}}=
\begin{pmatrix}
\Tilde{\lambda}_{1} & \Tilde{\lambda}_{1,2}\\
\Tilde{\lambda}_{1} & \Tilde{\lambda}_{2,2}\\
\Tilde{\lambda}_{1} & \Tilde{\lambda}_{3,2}\\
\end{pmatrix},
\label{eq:2f-gs}
\end{equation}
with \(i=1,2,3,\) and \(r=2\) factors.

\subsection{Application to the macro-financial dataset}

In the first part of the empirical analyses, we compare the score-driven factor model (Eq. \ref{eq:dfm:sd:obs} and \ref{eq:dfm:sd:trans}) with unrestricted loading matrix against the general lower-triangular (LT) restriction in Equation (\ref{eq:lt}) with respect to a set of \(N=8\) US macro and financial time series from January 1981 until August 2024 (\(T = 524\) months).\footnote{Given the dataset, the group-based restrictions in Equations \eqref{eq:gs-lt}, \eqref{eq:gs}, and \eqref{eq:2f-gs}, are excluded from this analysis.} 

\begin{figure}[h!]
\centering 
\includegraphics[width=0.8\textwidth]{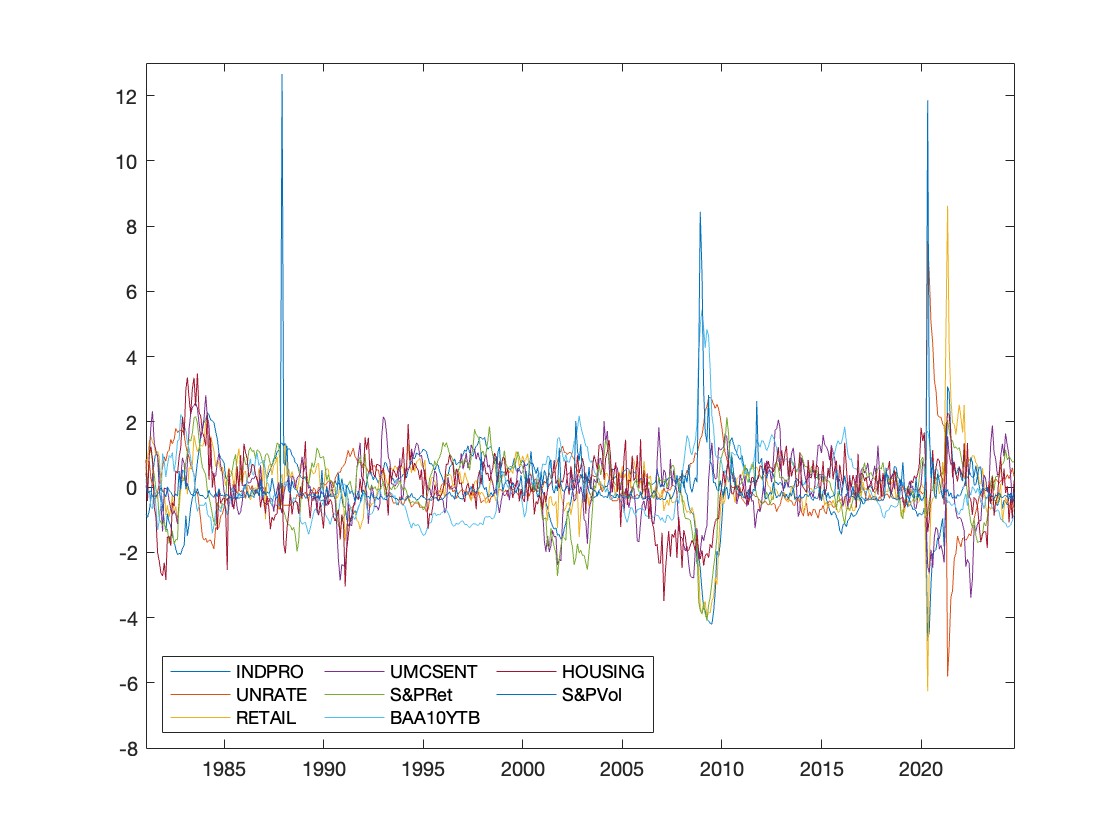}
\vspace{-0.5cm}
\caption{Macro and Financial Time Series from January 1981 to August 2024}
\label{fig:macro-fin_data}
\end{figure}

In particular, following \cite{creal2014observation}, we consider the annual change in log industrial production (INDPRO), the annual change in the unemployment rate (UNRATE), the annualized S\&P500 returns (S\&PRet) and volatility (S\&PVol), and the spread between the yield on Baa-rated bonds and the yield on 10-year Treasury bonds (BAA10YTB). In addition, the dataset contains the annual change in log retail sales (RETAIL), the annual change in the housing starts index (HOUSING), as well as the annual change in the survey-based consumer sentiment index (UMCSENT).\footnote{The data are retrieved from the FRED-MD macroeconomic database, while the S\&PVol series is the annualized monthly realized volatility of the S\&P500 Index constructed using the daily data from Yahoo Finance; see \cite{creal2014observation}.} 
We plot the standardized series in Figure \ref{fig:macro-fin_data}.

To perform the estimation, we consider the Student-$t$ innovations, allowing for distinct constant terms and \(A\) parameters that govern the dynamics of factors, while the coefficients of the matrix \(B\) are the same. 
We initialize the parameters based on the PCA estimates for loadings and respective group-based averages for correspondingly restricted matrices. For the parameters governing the dynamics of factors, we set  \(c\) and \(A\) equal to, respectively, the constant term and standard deviation estimates of an AR(1) model fitted to the principal components. Whereas, the initial value of the shared parameter \(B\) is set above 0.9 for all factors to reflect the persistence typically observed in the literature for such series.

\begin{table}[ht]
\begin{center}
\caption{In-Sample Fit} \label{table:1}
\begin{tabular}{l|c|c|c}
\hline
Model & LLF & AIC & BIC\\
\hline
\hline
1F Full & -3924.19 & \textit{15.05} & \textit{15.20}\\
\hline
1F LT & -3958.39 & 15.18 & 15.33\\
\hline
2F Full & -3368.10 & \textit{12.97} & \textit{13.20}\\
\hline
2F LT & -3451.66 & 13.28 & 13.50\\
\hline
3F Full & -2755.02 & \textbf{10.66} & \textbf{10.98}\\
\hline
3F LT & -2813.03 & 10.87 & 11.14\\
\hline
4F Full & -3497.67 & \textit{13.54} & \textit{13.93}\\
\hline
4F LT & -4642.37 & 17.87 & 18.20\\
\hline
\hline
\end{tabular}
\end{center}
\scriptsize{\textit{Notes}: This table reports the value of the maximized log-likelihood function (LLF), AIC, and BIC criteria of the unrestricted score-driven factor models with either 1, 2, 3, or 4 factors, as well as the models with the general LT restriction applied on the loading matrix. The values in bold correspond to the best model for each criterion. The values in italics correspond to the best fitting model for the corresponding number of factors. The models are estimated on the cross-section of \(N=8\) macroeconomic and financial variables from January 1981 until August 2024 (\(T=524\) months).} 
\end{table}

The in-sample results (Table \ref{table:1}) evaluated in terms of the value of the maximized log-likelihood function (LLF), the Akaike information criterion (AIC), and the Bayesian information criterion (BIC), suggest that the 3F Full model provides the best fit to the data in terms of both AIC and BIC. Regardless of the number of factors, the models with unconstrained loadings, i.e., the so-called Full models, always significantly outperform the corresponding models with LT restrictions. As follows, all the restricted models are rejected against the unrestricted ones, with the $p$-value of each LR statistics below \(0.01\); see Table \ref{table:2}.

\begin{figure}[h!]
\centering
\includegraphics[width=.8\linewidth]{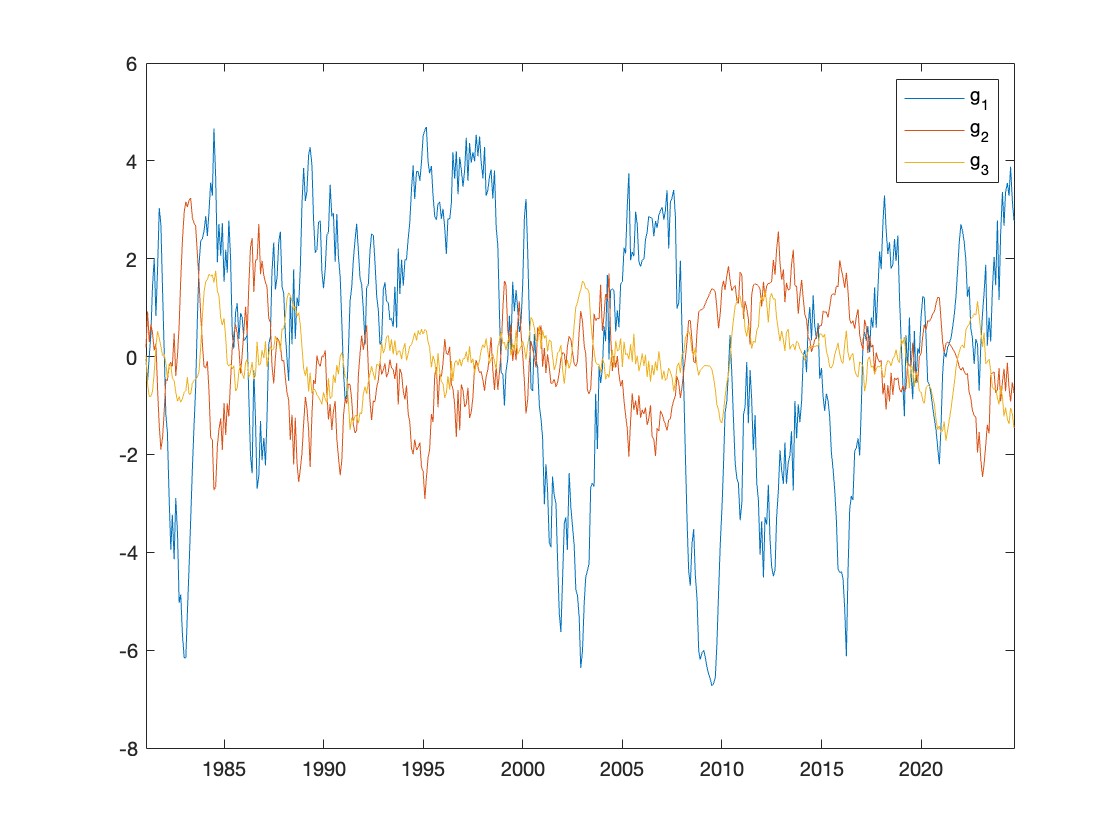}
\includegraphics[width=.8\linewidth]{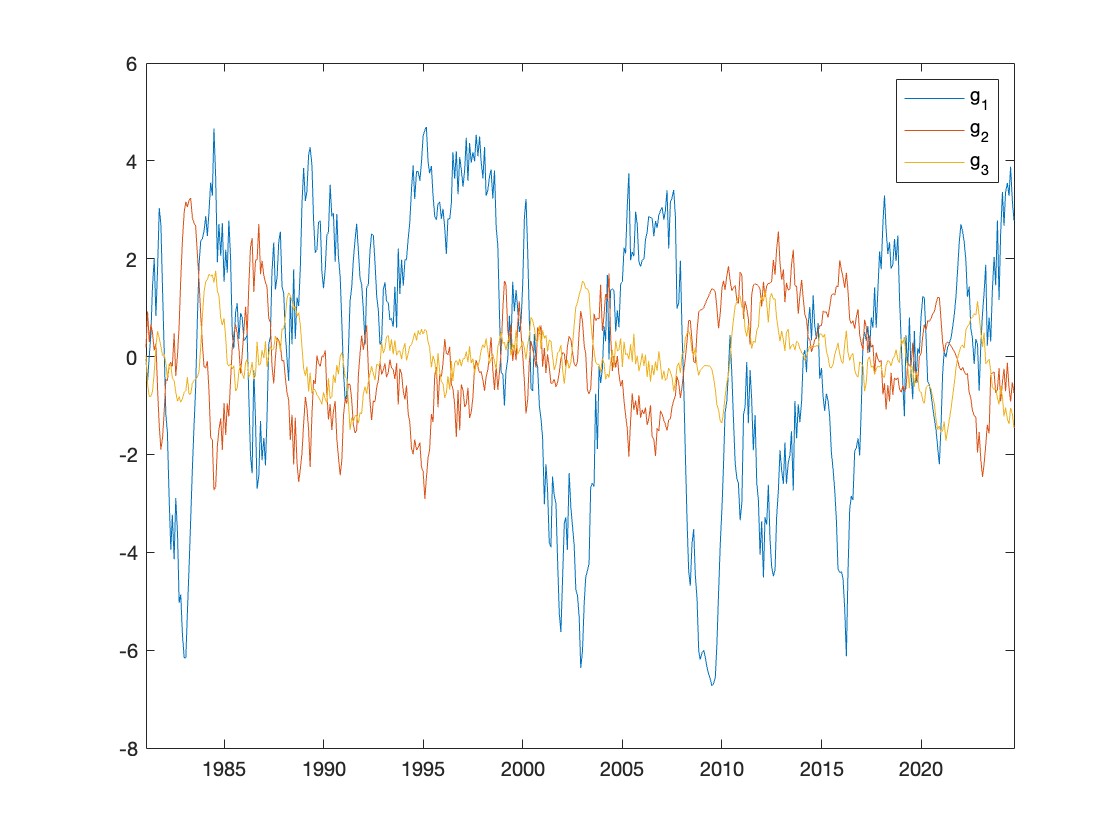}
\caption{3F Full fitted factors under shifted time series ordering}
\label{fig:full_order_shift}
\end{figure}

\begin{figure}[h!]
\centering
\includegraphics[width=.8\linewidth]{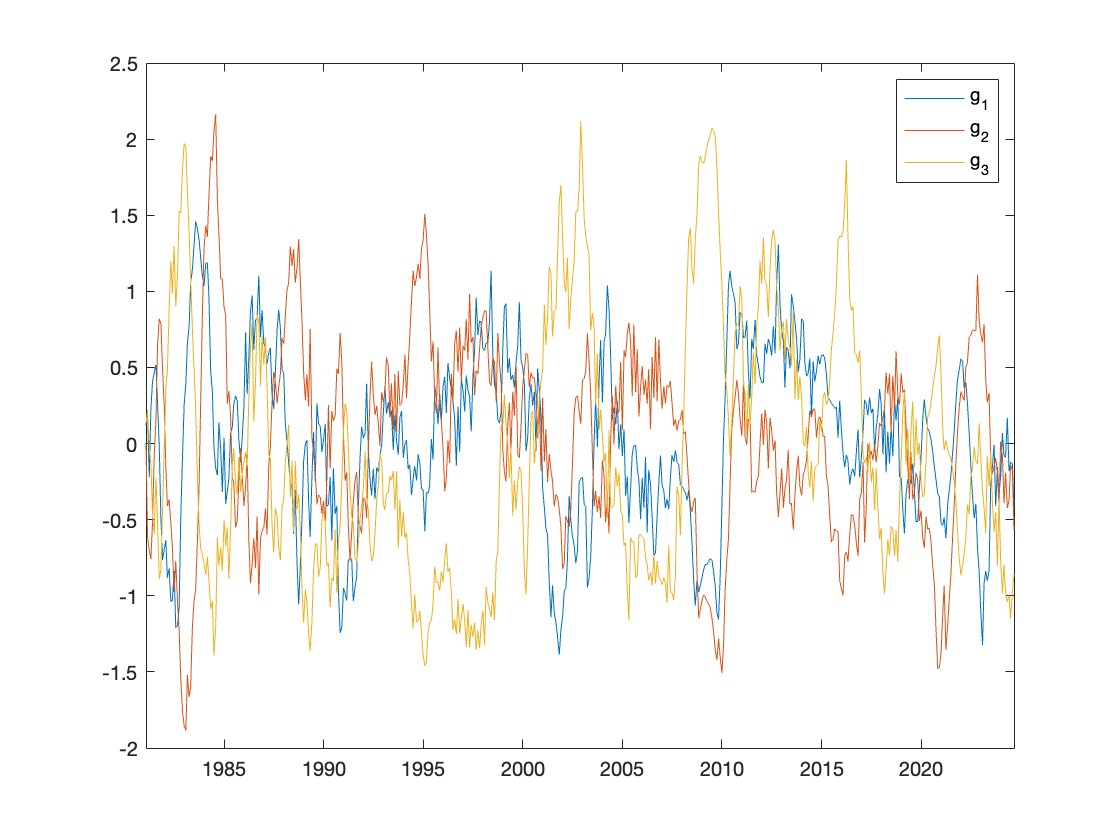}
\includegraphics[width=.8\linewidth]{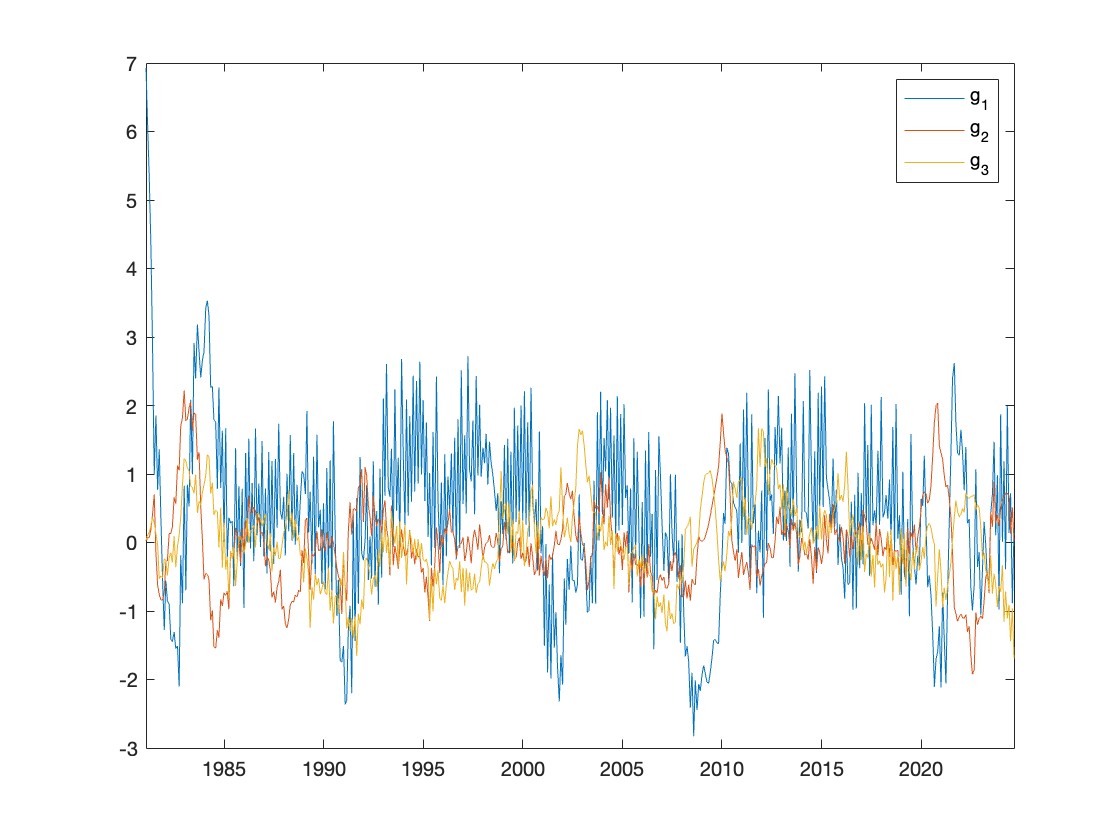}
\caption{3F LT fitted factors under shifted time series ordering}
\label{fig:lt_order_shift}
\end{figure}

Next, we assess the robustness of the estimated factor dynamics with respect to changes in the ordering of the observed variables. To this purpose, we examine the dynamics of the estimated factors of the best fitting model, 3F Full, and the corresponding restriction, 3F LT, under the shifted cross-sectional ordering of the data. Clearly, Figure \ref{fig:full_order_shift} confirms that the 3F Full fitted factors are order-invariant, whereas the law of motion of the 3F LT factors depends on the ordering (Figure \ref{fig:lt_order_shift}). Moreover, the first fitted factor of the 3F Full model (Figure \ref{fig:full_order_shift}) captures the business cycle dynamics, with its troughs aligning with periods of economic crises, such as the dot-com bubble in 2002, the 2007-2008 financial crisis, and the 2020 COVID pandemic.

\begin{table} [h!] 
\begin{center}
\caption{Likelihood Ratio Tests} \label{table:2}
\begin{tabular}{l|c|c|c}
\hline
Models & LR & df. & $p$-value\\
\hline
\hline
2F LT/2F Full & 167.12 & 2 &\(<\)0.01\\
\hline
3F LT/3F Full & 116.02 & 5 &\(<\)0.01\\
\hline
4F LT/4F Full & 2289.40 & 9 &\(<\)0.01\\
\hline
\hline
\end{tabular}
\end{center}
\scriptsize{\textit{Notes}: Likelihood ratio (LR) statistics, their degrees of freedom, and corresponding $p$-values for testing the general LT restricted models against the unrestricted ones. The models are estimated on the cross-section of \(N=8\) macroeconomic and financial variables from January 1981 until August 2024 (\(T=524\) months).}  
\end{table}

Finally, we perform a forecasting comparison of models based on their in-sample and out-of-sample MSE losses (Table \ref{table:3}). For the out-of-sample analysis, we use a rolling window approach, re-estimating the models over \(T_e\) = 312 months and generating one-month-ahead forecasts, for a total of 212 forecasts. The model rankings remain the same across both evaluations. Notably, the 4F Full model achieves the lowest average MSE loss, while each Full model outperforms its corresponding LT restriction.

\begin{table}[h!]
\begin{center}
\caption{In- and Out-of-Sample MSE} \label{table:3}
\begin{tabular}{l|c|c}
\hline
Model & In-Sample MSE& Out-of-Sample MSE\\
\hline
\hline
1F Full & \textit{0.740}& \textit{1.066}\\
\hline
1F LT & 0.755& 1.243\\
\hline
2F Full & \textit{0.621}& \textit{0.999}\\
\hline
2F LT & 0.715& 1.049\\
\hline
3F Full & \textit{0.586}& \textit{0.948}\\
\hline
3F LT & 0.598& 0.981\\
\hline
4F Full & \textbf{0.489}& \textbf{0.788}\\
\hline
4F LT & 0.789& 0.903\\
\hline
\hline
\end{tabular}
\end{center}
\scriptsize{\textit{Notes}: This table reports the average in- and out-of-sample MSE loss of the unrestricted score-driven factor models with either 1, 2, 3, or 4 factors, as well as the models with the general LT restriction applied on the loading matrix. The values in bold correspond to the best in-/out-of-sample model overall. The values in italics correspond to the best in-/out-of-sample model for the corresponding number of factors. The models are estimated on the cross-section of \(N=8\) macroeconomic and financial variables from January 1981 until August 2024 (\(T=524\) months), while the one-month-ahead out-of-sample forecasts are generated using a rolling-window scheme of \(T=312\) months (a total of 212 forecasts).} 
\end{table}

\clearpage
\subsection{Application to daily returns of S\&P500 constituents}

In the second part of the empirical application, we compare the score-driven factor model (Eq. \eqref{eq:dfm:sd:obs} and \eqref{eq:dfm:sd:trans}) with a full set of discussed specifications with restricted loading matrix, i.e., Equations \eqref{eq:lt}, \eqref{eq:gs-lt}, \eqref{eq:gs}, and \eqref{eq:2f-gs}. 

Except for the LT restriction in Equation \ref{eq:lt}, where the scale is identifiable, the parameter \(c_1\) is pre-set in the remaining restrictions in order to fix the scale of the model parameters.
We estimate the models on the cross-section of \(N=87\) open-to-close log-returns of the S\&P500 Index constituents over the period from January 2, 2001 until December 31, 2014 (\(T=3521\)) adopted by \cite{opschoor2021closed}. Table \ref{table:D1} in the Appendix provides a list of the ticker symbols and the corresponding industry classification for all the stocks.

The full set of candidate models includes the unrestricted score-driven factor models with 2, 5, and 11 factors (2F Full, 5F Full, and 11F Full, respectively), the 2 factor model with the common and group-specific loadings (2F GS; see \eqref{eq:2f-gs}), the 5 and 11 factor models with the general lower-triangular restriction (5F LT and 11F LT, respectively), coupled with the model with the group-specific lower-triangular loadings (10F GS-LT; see \eqref{eq:gs-lt}), and the specification with the common and group-specific factor loading matrix (11F GS; see \eqref{eq:gs}). The choice of the number of factors is driven by \(p=10\) groups available in the dataset, together with the specifications of the loadings restrictions. 


\begin{table} [h!] 
\begin{center}
\caption{Estimation Results} \label{table:4}
\begin{tabular}{l|c|c|c}
\hline
Model & LLF & AIC & BIC\\
\hline
\hline
2F Full & 61813.10 & \textit{-35.10} & \textit{-35.05}\\
\hline
2F GS & 61382.78 & -34.86 & -34.84\\
\hline
5F Full & 64594.04 & \textit{-36.66} & \textit{-36.57}\\
\hline
5F LT & 60867.16 & -34.54 & -34.45\\
\hline
11F Full & 70925.08 & \textbf{-40.23} & \textbf{-40.04}\\
\hline
11F LT & 70219.19 & -39.83 & -39.65\\
\hline
10F GS-LT & 68328.20 & -38.80 & -38.77\\
\hline
11F GS & 66834.39 & -37.96 & -37.94\\
\hline
\hline
\end{tabular}
\end{center}
\vspace{-0.4cm}
\scriptsize{\textit{Notes}: This table reports the value of the maximized log-likelihood function (LLF), AIC, and BIC criteria of the unrestricted score-driven factor models with either 2, 5, or 11 factors, as well as the models with distinct restrictions applied on the loading matrix. The values in bold correspond to the best model for each criterion. The values in italics correspond to the best fitting model for the corresponding number of factors.
The models are estimated on the cross-section of \(N=87\) S\&P500 constituents from January 2, 2001 until December 31, 2014 (\(T=3521\)).} 
\end{table}

Table \ref{table:4} summarizes the full-sample estimation results.\footnote{The parameters are initialized in the same manner as for the macro-financial dataset.} We derive several conclusions. First, the 11F Full model outperforms all the other models in terms of both AIC and BIC criteria. Second, the models with restricted loadings are always inferior compared to the Full model with the same number of factors, with the best AIC and BIC among the former achieved under the general lower-triangular restriction (i.e., LT). Hence, reducing the model complexity by grouping the data based on the observed characteristics, such as the industry classification, significantly reduces the model fitting.

These results are supperted by the likelihood ratio (LR) tests shown in Table \ref{table:5}.\ 
The table reports the LR statistics for testing the null hypothesis of a selected simpler version with restricted loadings against the unrestricted model that nests it. The $p$-value of each LR statistic is below 0.01, assuming a $\chi^2$ distribution with the reported number of degrees of freedom.

\begin{table} [h!] 
\begin{center}
\caption{Likelihood Ratio Tests } \label{table:5}
\begin{tabular}{l|c|c|c}
\hline
Models & LR & df. & $p$-value\\
\hline
\hline
2F GS/2F Full & 860.65 & 163 &\(<\)0.01\\
\hline
5F LT/5F Full & 7453.77 & 14 &\(<\)0.01\\
\hline
11F LT/11F Full & 1411.79 & 65 &\(<\)0.01\\
\hline
10F GS-LT/11F Full & 5193.76 & 900 &\(<\)0.01\\
\hline
11F GS/11F Full & 8181.38 & 946 &\(<\)0.01\\
\hline
\hline
\end{tabular}
\end{center}
\vspace{-0.4cm}
\scriptsize{\textit{Notes}: Likelihood ratio (LR) statistics, their degrees of freedom, and corresponding $p$-values for testing the selected restricted models against the unrestricted ones. The models are estimated on the cross-section of \(N=87\) S\&P500 constituents from January 2, 2001 until December 31, 2014 (\(T=3521\)).} 
\end{table}

Given that the assumption of constant loadings may be considered restrictive, we perform the full in-sample analyses for the analogous set of models with time-varying loading dynamics in Equations \eqref{eq:dfm:sd:obs_tv}-\eqref{eq:gas_g}. In particular, to avoid the parameter proliferation, we assume a scalar structure for \(\bm{A}^{(l)}\) and \(\bm{B}^{(l)}\), and target the unconditional mean \(\mathbb{E}[\bm{l}_t]\) by utilizing the corresponding static estimate of $\bm{\Lambda}$. As such, all the dynamic models have two additional parameters with respect to the static versions. 
The corresponding estimation results and LR tests are presented in Tables \ref{table:6} and \ref{table:7}.

The comparison of the results in Tables \ref{table:4} and \ref{table:6} indicates that each dynamic model significantly outperforms the corresponding static version. Regarding the former, we confirm that the Full model with either 2, 5, or 11 factors is superior to the corresponding restricted versions in terms of both AIC and BIC. Furthermore, the dynamic 11F model with the general lower-triangular restriction, i.e., 11F LT, prevails over the models with group-based loading matrices, whereas the best-fitting model for each information criterion is the 11F Full model. Ultimately, the LR test results reported in Table \ref{table:7} show that all the restricted dynamic models are rejected against the corresponding unrestricted version at the 1\% significance level.

\begin{table}[h!]
\begin{center}
\caption{Estimation Results with time-varying loadings} \label{table:6}
\begin{tabular}{l|c|c|c}
\hline
Model & LLF & AIC & BIC\\
\hline
\hline
2F Full & 61917.76 & \textit{-35.16} & \textit{-35.11}\\
\hline
2F GS & 61412.40 & -34.88 & -34.86\\
\hline
5F Full & 130778.83 & \textit{-74.26} & \textit{-74.16}\\
\hline
5F LT & 78395.97 & -44.50 & -44.41\\
\hline
11F Full & 159676.71 & \textbf{-90.64} & \textbf{-90.45}\\
\hline
11F LT & 137712.14 & -78.17 & -77.99\\
\hline
10F GS-LT & 97347.46 & -55.29 & -55.26\\
\hline
11F GS & 98769.97 & -56.10 & -56.08\\
\hline
\hline
\end{tabular}
\end{center}
\vspace{-0.4cm}
\scriptsize{\textit{Notes}: This table reports the value of the maximized log-likelihood function (LLF), AIC, and BIC criteria of the unrestricted score-driven factor models with either 2, 5, or 11 factors and dynamic loadings, as well as the models with distinct restrictions applied on the time-varying loading matrix. The values in bold correspond to the best model for each criterion. The values in italics correspond to the best fitting model for the corresponding number of factors.
The models are estimated on the cross-section of \(N=87\) S\&P500 constituents from January 2, 2001 until December 31, 2014 (\(T=3521\)).} 
\end{table}

\begin{table} [h!] 
\begin{center}
\caption{Likelihood Ratio Tests for models with time-varying loadings} \label{table:7}
\begin{tabular}{l|c|c|c}
\hline
Models & LR & df. & $p$-value\\
\hline
\hline
2F GS/2F Full & 1010.72 & 163 &\(<\)0.01\\
\hline
5F LT/5F Full & 104765.72 & 14 &\(<\)0.01\\
\hline
11F LT/11F Full & 43929.14 & 65 &\(<\)0.01\\
\hline
10F GS-LT/11F Full & 124658.50 & 900 &\(<\)0.01\\
\hline
11F GS/11F Full & 121813.48 & 946 &\(<\)0.01\\
\hline
\hline
\end{tabular}
\end{center}
\vspace{-0.4cm}
\scriptsize{\textit{Notes}: Likelihood ratio (LR) statistics, their degrees of freedom, and corresponding $p$-values for testing the selected restricted models with dynamic loadings against the unrestricted ones. The models are estimated on the cross-section of \(N=87\) S\&P500 constituents from January 2, 2001 until December 31, 2014 (\(T=3521\)).} 
\end{table}

\begin{figure}[h!]
\centering 
\includegraphics[width=0.8\textwidth]{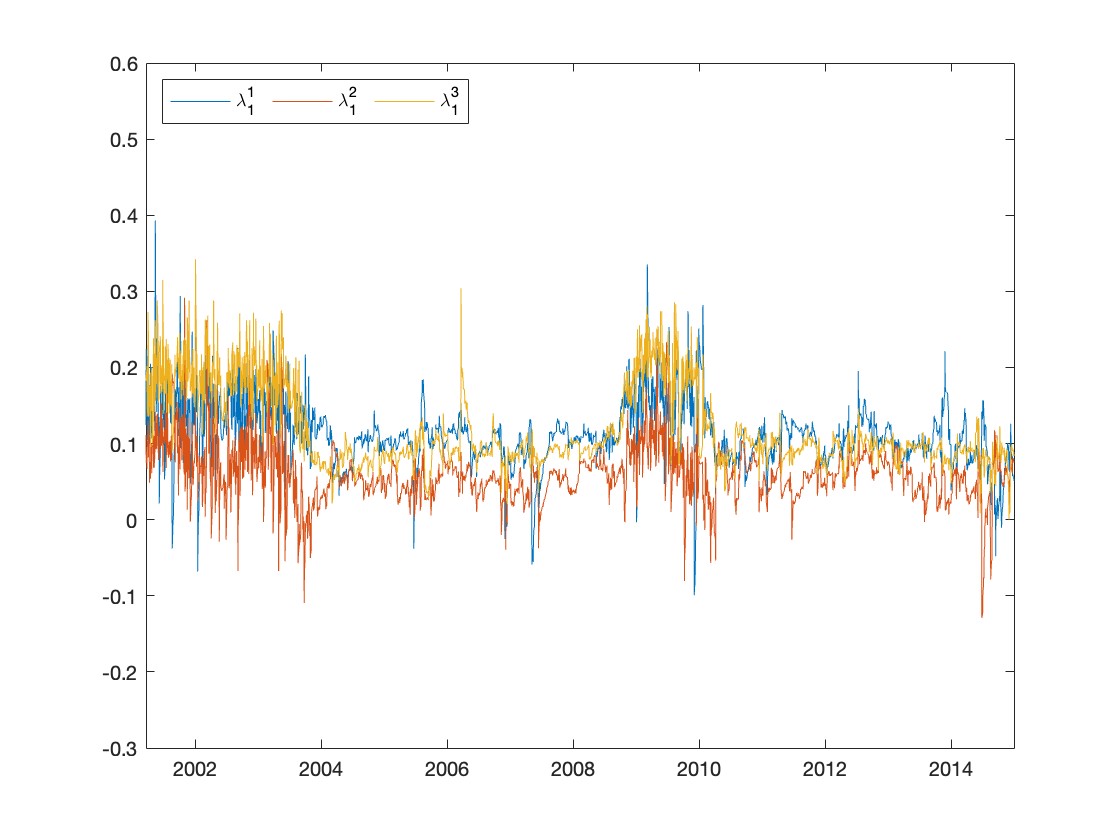}
\vspace{-0.5cm}
\caption{11F Full fitted dynamic loadings on the first factor for AA, BA, and CAT}
\label{fig:dyn.load}
\end{figure}

In Figure \ref{fig:dyn.load}, we show an example of the fitted dynamic loadings. The loading paths exhibit notable shifts during the periods of financial and/or economic turbulence, e.g., the dot-com bubble, the Great Recession, etc., confirming that the assumption of the constant relationship between the returns of the selected S\&P500 constituents and the underlying factors may be too restrictive.

We conclude that the flexibility of the score-driven framework under the introduced identification scheme significantly increases the model ability to capture the dependence dynamics of both financial and macroeconomic time series since the models with unrestricted loadings always outperform each of the restricted specifications in a highly statistically significant way.
\newpage
\section{Conclusions}
\label{sec:concl}

This paper studies the identifiability of score-driven factor models, demonstrating that, under fairly general assumptions on the parameter space, the static and time-varying parameters can be determined up to multiplicative scalar constant. 
We further extend our results to include score-driven factor models with dynamic loadings and nonlinear factor models. 
This property distinguishes score-driven models from parameter-driven factor models, which are only identifiable up to rotations.\ The ability to identify score-driven models more precisely under minimal restrictions enhances their economic interpretability.


Our theoretical findings are validated through extensive simulations, confirming the practical applicability and robustness of the proposed methods. The results highlight the advantages of score-driven factor models in terms of both identifiability and ease of estimation, making them a valuable tool for economic and financial time series analysis.

In the empirical application, we verify that the score-driven framework under the introduced flexible identification scheme provides for significant in- and out-of-sample gains with respect to the typically applied restrictions in modelling both the financial and macroeconomic time series dynamics.

\clearpage
\bibliographystyle{elsarticle-harv}
\bibliography{paper.bbl}

\begin{thebibliography}{24}
\expandafter\ifx\csname natexlab\endcsname\relax\def\natexlab#1{#1}\fi
\providecommand{\url}[1]{\texttt{#1}}
\providecommand{\href}[2]{#2}
\providecommand{\path}[1]{#1}
\providecommand{\DOIprefix}{doi:}
\providecommand{\ArXivprefix}{arXiv:}
\providecommand{\URLprefix}{URL: }
\providecommand{\Pubmedprefix}{pmid:}
\providecommand{\doi}[1]{\href{http://dx.doi.org/#1}{\path{#1}}}
\providecommand{\Pubmed}[1]{\href{pmid:#1}{\path{#1}}}
\providecommand{\bibinfo}[2]{#2}
\ifx\xfnm\relax \def\xfnm[#1]{\unskip,\space#1}\fi
\bibitem[{Adrian and Franzoni(2009)}]{adrian2009learning}
\bibinfo{author}{Adrian, T.}, \bibinfo{author}{Franzoni, F.}, \bibinfo{year}{2009}.
\newblock \bibinfo{title}{Learning about beta: Time-varying factor loadings, expected returns, and the conditional capm}.
\newblock \bibinfo{journal}{Journal of Empirical Finance} \bibinfo{volume}{16}, \bibinfo{pages}{537--556}.
\bibitem[{Artemova(2023)}]{artemova}
\bibinfo{author}{Artemova, M.}, \bibinfo{year}{2023}.
\newblock \bibinfo{title}{An Order-Invariant Score-Driven Dynamic Factor Model}.
\newblock \bibinfo{type}{Working paper}.
\bibitem[{Artemova et~al.(2022)Artemova, Blasques, van Brummelen and Koopman}]{artemova2022score}
\bibinfo{author}{Artemova, M.}, \bibinfo{author}{Blasques, F.}, \bibinfo{author}{van Brummelen, J.}, \bibinfo{author}{Koopman, S.J.}, \bibinfo{year}{2022}.
\newblock \bibinfo{title}{Score-driven models: Methods and applications}, in: \bibinfo{booktitle}{Oxford Research Encyclopedia of Economics and Finance}. \bibinfo{publisher}{Oxford University Press}.
\bibitem[{Bai and Li(2012)}]{bai2012statistical}
\bibinfo{author}{Bai, J.}, \bibinfo{author}{Li, K.}, \bibinfo{year}{2012}.
\newblock \bibinfo{title}{Statistical analysis of factor models of high dimension}.
\newblock \bibinfo{journal}{The Annals of Statistics} , \bibinfo{pages}{436 -- 465}.
\bibitem[{Blasques et~al.(2024)Blasques, van Brummelen, Gorgi and Koopman}]{blasques2024maximum}
\bibinfo{author}{Blasques, F.}, \bibinfo{author}{van Brummelen, J.}, \bibinfo{author}{Gorgi, P.}, \bibinfo{author}{Koopman, S.J.}, \bibinfo{year}{2024}.
\newblock \bibinfo{title}{Maximum likelihood estimation for non-stationary location models with mixture of normal distributions}.
\newblock \bibinfo{journal}{Journal of Econometrics} \bibinfo{volume}{238}, \bibinfo{pages}{105575}.
\bibitem[{Chamberlain(1983)}]{chamberlain1983funds}
\bibinfo{author}{Chamberlain, G.}, \bibinfo{year}{1983}.
\newblock \bibinfo{title}{Funds, factors, and diversification in arbitrage pricing models}.
\newblock \bibinfo{journal}{Econometrica} , \bibinfo{pages}{1305--1323}.
\bibitem[{Cox(1981)}]{Cox}
\bibinfo{author}{Cox, D.}, \bibinfo{year}{1981}.
\newblock \bibinfo{title}{Statistical analysis of time series: Some recent developments [with discussion and reply]}.
\newblock \bibinfo{journal}{Scandinavian Journal of Statistics} \bibinfo{volume}{8}, \bibinfo{pages}{93--115}.
\bibitem[{Creal et~al.(2011)Creal, Koopman and Lucas}]{creal2011dynamic}
\bibinfo{author}{Creal, D.}, \bibinfo{author}{Koopman, S.J.}, \bibinfo{author}{Lucas, A.}, \bibinfo{year}{2011}.
\newblock \bibinfo{title}{A dynamic multivariate heavy-tailed model for time-varying volatilities and correlations}.
\newblock \bibinfo{journal}{Journal of Business \& Economic Statistics} \bibinfo{volume}{29}, \bibinfo{pages}{552--563}.
\bibitem[{Creal et~al.(2013)Creal, Koopman and Lucas}]{creal2013generalized}
\bibinfo{author}{Creal, D.}, \bibinfo{author}{Koopman, S.J.}, \bibinfo{author}{Lucas, A.}, \bibinfo{year}{2013}.
\newblock \bibinfo{title}{Generalized autoregressive score models with applications}.
\newblock \bibinfo{journal}{Journal of {A}pplied {E}conometrics} \bibinfo{volume}{28}, \bibinfo{pages}{777--795}.
\bibitem[{Creal et~al.(2014)Creal, Schwaab, Koopman and Lucas}]{creal2014observation}
\bibinfo{author}{Creal, D.}, \bibinfo{author}{Schwaab, B.}, \bibinfo{author}{Koopman, S.J.}, \bibinfo{author}{Lucas, A.}, \bibinfo{year}{2014}.
\newblock \bibinfo{title}{Observation-driven mixed-measurement dynamic factor models with an application to credit risk}.
\newblock \bibinfo{journal}{Review of Economics and Statistics} \bibinfo{volume}{96}, \bibinfo{pages}{898--915}.
\bibitem[{Doz et~al.(2012)Doz, Giannone and Reichlin}]{doz2012quasi}
\bibinfo{author}{Doz, C.}, \bibinfo{author}{Giannone, D.}, \bibinfo{author}{Reichlin, L.}, \bibinfo{year}{2012}.
\newblock \bibinfo{title}{A quasi--maximum likelihood approach for large, approximate dynamic factor models}.
\newblock \bibinfo{journal}{Review of economics and statistics} \bibinfo{volume}{94}, \bibinfo{pages}{1014--1024}.
\bibitem[{D’Innocenzo and Lucas(2024)}]{d2024dynamic}
\bibinfo{author}{D’Innocenzo, E.}, \bibinfo{author}{Lucas, A.}, \bibinfo{year}{2024}.
\newblock \bibinfo{title}{Dynamic partial correlation models}.
\newblock \bibinfo{journal}{Journal of Econometrics} \bibinfo{volume}{241}, \bibinfo{pages}{105747}.
\bibitem[{Geweke(1977)}]{geweke1977dynamic}
\bibinfo{author}{Geweke, J.}, \bibinfo{year}{1977}.
\newblock \bibinfo{title}{The dynamic factor analysis of economic time series}.
\newblock \bibinfo{journal}{Latent variables in socio-economic models} , \bibinfo{pages}{365--383}.
\bibitem[{Giglio et~al.(2022)Giglio, Kelly and Xiu}]{giglio2022factor}
\bibinfo{author}{Giglio, S.}, \bibinfo{author}{Kelly, B.}, \bibinfo{author}{Xiu, D.}, \bibinfo{year}{2022}.
\newblock \bibinfo{title}{Factor models, machine learning, and asset pricing}.
\newblock \bibinfo{journal}{Annual Review of Financial Economics} \bibinfo{volume}{14}, \bibinfo{pages}{337--368}.
\bibitem[{Harvey(1990)}]{harvey_1990}
\bibinfo{author}{Harvey, A.C.}, \bibinfo{year}{1990}.
\newblock \bibinfo{title}{Forecasting, Structural Time Series Models and the Kalman Filter}.
\newblock \bibinfo{publisher}{Cambridge University Press}.
\bibitem[{Harvey(2013)}]{harvey2013dynamic}
\bibinfo{author}{Harvey, A.C.}, \bibinfo{year}{2013}.
\newblock \bibinfo{title}{Dynamic models for volatility and heavy tails: with applications to financial and economic time series}. volume~\bibinfo{volume}{52}.
\newblock \bibinfo{publisher}{Cambridge University Press}.
\bibitem[{Hillebrand et~al.(2023)Hillebrand, Mikkelsen, Spreng and Urga}]{hillebrand2023exchange}
\bibinfo{author}{Hillebrand, E.}, \bibinfo{author}{Mikkelsen, J.G.}, \bibinfo{author}{Spreng, L.}, \bibinfo{author}{Urga, G.}, \bibinfo{year}{2023}.
\newblock \bibinfo{title}{Exchange rates and macroeconomic fundamentals: Evidence of instabilities from time-varying factor loadings}.
\newblock \bibinfo{journal}{Journal of Applied Econometrics} \bibinfo{volume}{38}, \bibinfo{pages}{857--877}.
\bibitem[{Kelly et~al.(2019)Kelly, Pruitt and Su}]{kelly2019characteristics}
\bibinfo{author}{Kelly, B.T.}, \bibinfo{author}{Pruitt, S.}, \bibinfo{author}{Su, Y.}, \bibinfo{year}{2019}.
\newblock \bibinfo{title}{Characteristics are covariances: A unified model of risk and return}.
\newblock \bibinfo{journal}{Journal of Financial Economics} \bibinfo{volume}{134}, \bibinfo{pages}{501--524}.
\bibitem[{Mikkelsen et~al.(2019)Mikkelsen, Hillebrand and Urga}]{mikkelsen2019consistent}
\bibinfo{author}{Mikkelsen, J.G.}, \bibinfo{author}{Hillebrand, E.}, \bibinfo{author}{Urga, G.}, \bibinfo{year}{2019}.
\newblock \bibinfo{title}{Consistent estimation of time-varying loadings in high-dimensional factor models}.
\newblock \bibinfo{journal}{Journal of Econometrics} \bibinfo{volume}{208}, \bibinfo{pages}{535--562}.
\bibitem[{Oh and Patton(2018)}]{oh2018time}
\bibinfo{author}{Oh, D.H.}, \bibinfo{author}{Patton, A.J.}, \bibinfo{year}{2018}.
\newblock \bibinfo{title}{Time-varying systemic risk: Evidence from a dynamic copula model of cds spreads}.
\newblock \bibinfo{journal}{Journal of Business \& Economic Statistics} \bibinfo{volume}{36}, \bibinfo{pages}{181--195}.
\bibitem[{Opschoor et~al.(2021)Opschoor, Lucas, Barra and Van~Dijk}]{opschoor2021closed}
\bibinfo{author}{Opschoor, A.}, \bibinfo{author}{Lucas, A.}, \bibinfo{author}{Barra, I.}, \bibinfo{author}{Van~Dijk, D.}, \bibinfo{year}{2021}.
\newblock \bibinfo{title}{Closed-form multi-factor copula models with observation-driven dynamic factor loadings}.
\newblock \bibinfo{journal}{Journal of Business \& Economic Statistics} \bibinfo{volume}{39}, \bibinfo{pages}{1066--1079}.
\bibitem[{Opschoor et~al.(2024)Opschoor, Lucas and Rossini}]{opschoor2024conditional}
\bibinfo{author}{Opschoor, A.}, \bibinfo{author}{Lucas, A.}, \bibinfo{author}{Rossini, L.}, \bibinfo{year}{2024}.
\newblock \bibinfo{title}{The conditional autoregressive f-riesz model for realized covariance matrices}.
\newblock \bibinfo{journal}{Journal of Financial Econometrics} , \bibinfo{pages}{nbae023}.
\bibitem[{Stock and Watson(2011)}]{StockWatson2011}
\bibinfo{author}{Stock, J.H.}, \bibinfo{author}{Watson, M.W.}, \bibinfo{year}{2011}.
\newblock \bibinfo{title}{35 dynamic factor models}, in: \bibinfo{booktitle}{The Oxford Handbook of Economic Forecasting}. \bibinfo{publisher}{Oxford University Press}.
\bibitem[{Xu(2022)}]{xu2022testing}
\bibinfo{author}{Xu, W.}, \bibinfo{year}{2022}.
\newblock \bibinfo{title}{Testing for time-varying factor loadings in high-dimensional factor models}.
\newblock \bibinfo{journal}{Econometric Reviews} \bibinfo{volume}{41}, \bibinfo{pages}{918--965}.

\end{thebibliography}

\newpage

\appendix

\bigskip
\begin{center}
{\LARGE\bf Appendix}
\end{center}

\section{Proofs of the main results}

\subsection{Proof of Proposition \textbf{\eqref{prop:AffineScore}}}
\label{app:proof}
We can re-write Equations \eqref{eq:dfm:sd:obs}, \eqref{eq:dfm:sd:trans} as follows
\begin{align}
\bm{y}_t &= \overline{\bm{\Lambda}}\ \overline{\bm{f}}_t+\bm{\epsilon}_t \\
\overline{\bm{f}}_{t+1} &= \overline{\bm{c}} + \bm{T}^{-1}\bm{A}\bm{s}_t + \overline{\bm{B}}\ \overline{\bm{f}}_t\label{eq:app:sd:trans}
\end{align}
Let us observe that, using the chain rule, the score in Equation \eqref{eq:score} can be re-written as
\begin{equation}
\bm{\nabla}_t = \left[\frac{\partial\log p(\bm{y}_t|\bm{\mathcal{F}}_{t-1})}{\partial \bm{f}_t}\right]'=\left[\frac{\partial\log p(\bm{y}_t|\bm{f}_t)}{\partial\overline{\bm{f}}_t}\frac{\partial\overline{\bm{f}}}{\partial\bm{f}_t}\right]' = \bm{T}^{-1\prime}\overline{\bm{\nabla}}_t
\end{equation} 
Similarly, the conditional Fisher Information matrix can be written as follows
\begin{equation}
\bm{\mathcal{I}}_{t|t-1}=\mathbb{E}[\bm{\nabla}_t\bm{\nabla}_t'|\bm{\mathcal{F}_{t-1}}] = \bm{T}^{-1\prime}\overline{\bm{\mathcal{I}}}_{t|t-1}\bm{T}^{-1}
\end{equation}
Therefore, we have:
\begin{align}
\bm{s}_t =\left(\bm{\mathcal{I}}_{t|t-1}\right)^{-\beta}\bm{\nabla}_t=\bm{T}^{\beta}(\overline{\bm{\mathcal{I}}}_{t|t-1})^{-\beta}(\bm{T}^{-1+\beta})'\overline{\bm{\nabla}}_t
\end{align}
Therefore, we can re-write Equation \eqref{eq:app:sd:trans} as
\begin{equation}
\overline{\bm{f}}_{t+1} = \overline{\bm{c}} + \overline{\bm{A}}\ \overline{\bm{s}}_t + \overline{\bm{B}}\ \overline{\bm{f}}_t
\end{equation}
where $\overline{\bm{A}}=\bm{T}^{-1}\bm{A}\bm{T}^{\beta}$ and $\overline{\bm{s}}_t = (\overline{\bm{\mathcal{I}}}_{t|t-1})^{-\beta}(\bm{T}^{-1+\beta})'\overline{\bm{\nabla}}_t$.

\subsection{Proof of Theorem \eqref{thm:commut}}
Clearly, if $\bm{T}$ is scalar, $(\bm{T}^{-1+\beta})'$  will commute with $(\overline{\bm{\mathcal{I}}}_{t|t-1})^{-\beta}$.\ Let us now prove the opposite implication.
Since $\overline{\bm{A}}=\bm{T}^{-1}\bm{A} \bm{T}^{\beta}$, restricting $\bm{A}$ to be diagonal with non-zero entries and keeping the same restriction for $\overline{\bm{A}}$ requires $\bm{T}$, and thus $(\bm{T}^{-1+\beta})'$, to likewise be diagonal with non-zero entries.\ Note also that, since $\bm{\mathcal{I}}_{t|t-1}$ is not block diagonal by Assumption \eqref{ass:I}, and $\bm{T}$ is diagonal with non-zero entries, also $\overline{\bm{\mathcal{I}}}_{t|t-1}=\bm{T}'\bm{\mathcal{I}}_{t|t-1}\bm{T}$ is not block diagonal.\ Consequently, $(\overline{\bm{\mathcal{I}}}_{t|t-1})^{-\beta}$ is not block diagonal.\  

Let us consider the case in which the diagonal entries of $(\bm{T}^{-1+\beta})'$ are distinct and then the case in which there exist repeated entries.
If the entries of $(\bm{T}^{-1+\beta})'$ are distinct, it only commutes with diagonal matrices.\ However,  $(\overline{\bm{\mathcal{I}}}_{t|t-1})^{-\beta}$, being not block diagonal, cannot be diagonal.
In the second case, if some of the diagonal entries, say $k<r$, of $(\bm{T}^{-1+\beta})'$ are equal, it commutes with diagonal block matrices having a $k\times k$ block.\ However, this is still ruled out by the fact that $(\overline{\bm{\mathcal{I}}}_{t|t-1})^{-\beta}$ is not block diagonal.\ Therefore, the only case where  $(\overline{\bm{\mathcal{I}}}_{t|t-1})^{-\beta}$ and  $(\bm{T}^{-1+\beta})'$ commute is when $\bm{T}$ has $k=r$ equal diagonal entries, i.e., when $\bm{T}$ is scalar.

\subsection{Proof of Proposition \eqref{prop:order}}

Since $\bm{\Lambda}$ is unconstrained, the model for the permuted series can be written as $\tilde{\bm{y}}_t = \tilde{\bm{\Lambda}}\bm{f}_t + \tilde{\bm{\epsilon}}$, where $\tilde{\bm{y}}_t\coloneqq \bm{P}\bm{y}_t$, $\tilde{\bm{\Lambda}}\coloneqq\bm{P}\bm{\Lambda}$ is the new factor loading matrix and $\tilde{\bm{\epsilon}}\coloneqq\bm{P}\bm{\epsilon}$.\  The conditional likelihood of the permuted observations is 
\begin{align*}
    p(\tilde{\bm{y}}_t|\bm{\mathcal{F}}_{t-1},\bm{f}_t)&=\det{(\bm{P}\bm{\Sigma}\bm{P}')}^{-\frac{1}{2}}\psi ((\bm{P}\bm{y}_t-\bm{P}\bm{\Lambda}\bm{f}_t)'(\bm{P}\bm{\Sigma}\bm{P}')^{-1}(\bm{P}\bm{y}_t-\bm{P}\bm{\Lambda}\bm{f}_t))\\
    &= \det{(\bm{P}')}^{-\frac{1}{2}}\det{\bm{\Sigma}}^{-\frac{1}{2}}\det{\bm{P}}^{-\frac{1}{2}} \psi ((\bm{y}_t-\bm{\Lambda}\bm{f}_t)'\bm{P}'(\bm{P}')^{-1}\bm{\Sigma}^{-1}\bm{P}^{-1}\bm{P}(\bm{y}_t-\bm{\Lambda}\bm{f}_t))\\
    &= \det{\bm{\Sigma}}^{-\frac{1}{2}}\psi ((\bm{y}_t-\bm{\Lambda}\bm{f}_t)'\bm{\Sigma}^{-1}(\bm{y}_t-\bm{\Lambda}\bm{f}_t)),
\end{align*}
where the last line follows from the fact that the determinant of any permutation matrix is equal to 1 or $-1$, and that $\det{\bm{P}}=\det{\bm{P}'}$.\ Since the likelihood is invariant, the score $\bm{\nabla}_t$ and the Fisher information matrix $\bm{\mathcal{I}}_{t|t-1}$ are not affected by the permutation.\    Consequently, the filter dynamics are unchanged if the same starting value is used to initialize the filter.

\subsection{Proof of Proposition \eqref{prop:AffineScore_tv}}
Let $\bm{T}\in\mathbb{R}^{r\times r}$ be a non-singular matrix.\ The model in Equation \eqref{eq:dfm:sd:obs_tv} can be re-written as
\begin{equation}
\bm{y}_t = \bm{\Lambda}_t\bm{T}\bm{T}^{-1} \bm{g}_t + \bm{\epsilon}_t
\end{equation}
Let us define $\bm{\overline{\Lambda}}_t=\bm{\Lambda}_t\bm{T}$, $\bm{\overline{l}}_t=\text{vec}(\bm{\overline{\Lambda}}_t)$ and $\bm{\overline{g}}_t=\bm{T}^{-1}\bm{g}_t$.\ Since $\bm{\overline{\Lambda}}_t=\bm{\Lambda}_t\bm{T}=\bm{I}\bm{\Lambda}_t\bm{T}$, we have $\bm{\overline{l}}_t=\text{vec}(\bm{\overline{\Lambda}}_t)= \text{vec}(\bm{I}\bm{\Lambda}_t\bm{T})=(\bm{T}'\otimes \bm{I})\text{vec}(\bm{\Lambda}_t)$.\ In the last step, we have used the property $\text{vec}(\bm{PQR})=(\bm{R}'\otimes \bm{P})\text{vec}(\bm{Q})$, where $\otimes$ denotes the Kronecker product and $\bm{P}$, $\bm{Q}$, $\bm{R}$ are matrices of appropriate sizes.\ Therefore, it is possible to write Equation \eqref{eq:gas_l} as\medskip
\begin{equation}
\overline{\bm{l}}_{t+1}= \overline{\bm{c}}^{(l)}+ (\bm{T}'\otimes \bm{I})\bm{A}^{(l)} \bm{s}_t^{(l)} + \overline{\bm{B}}^{(l)} \overline{\bm{l}}_t\medskip
\label{eq:gas_l_partial}
\end{equation}
where $\overline{\bm{c}}^{(l)}=(\bm{T}'\otimes \bm{I})\bm{c}^{(l)}$ and $\overline{\bm{B}}^{(l)}=(\bm{T}'\otimes \bm{I})\bm{B}^{(l)}(\bm{T}'\otimes \bm{I})^{-1}$. Observe now that, using the chain rule, we can re-write the score $\bm{\nabla}_t^{(l)}$ as\medskip 
\begin{equation}
\bm{\nabla}_t^{(l)}=\left[\frac{\partial\log\mathcal{L}(\bm{y}_t|\bm{\mathcal{F}}_{t-1},\bm{l}_t,\bm{g}_t;\bm{\Theta}_{\bm{\epsilon}})}{\partial\bm{l}_t}\right]'=\left[\frac{\partial\log\mathcal{L}(\bm{y}_t|\bm{\mathcal{F}}_{t-1},\bm{l}_t,\bm{g}_t;\bm{\Theta}_{\bm{\epsilon}})}{\partial\bm{\overline{l}}_t}\times\frac{\partial\bm{\overline{l}}_t}{\partial\bm{l}_t }\right]'\medskip
\end{equation}
Since $\bm{\overline{l}}_t=\text{vec}(\bm{\overline{\Lambda}}_t)=(\bm{T}'\otimes \bm{I})\bm{l}_t$, we have $\frac{\partial\bm{\overline{l}}_t}{\partial\bm{l}_t }=\bm{T}'\otimes \bm{I}$. Thus\medskip
\begin{equation}
\bm{\nabla}_t^{(l)} = (\bm{T}\otimes\bm{I})\overline{\bm{\nabla}}_t^{(l)}\medskip
\end{equation}
where $\overline{\bm{\nabla}}_t^{(l)}=\left[\frac{\partial\log\mathcal{L}(\bm{y}_t|\bm{\mathcal{F}}_{t-1},\bm{l}_t,\bm{g}_t;\bm{\Theta}_{\bm{\epsilon}})}{\partial\bm{\overline{l}}_t}\right]'$.\ The conditional Fisher information can instead be written as\medskip
\begin{equation}
\bm{\mathcal{I}}_{t|t-1}^{(l)} = \mathbb{E}[\bm{\nabla}_t^{(l)}\bm{\nabla}_t^{(l)\prime}|\bm{\mathcal{F}}_{t-1}]=(\bm{T}\otimes\bm{I})\overline{\bm{\mathcal{I}}}_{t|t-1}^{(l)}(\bm{T}'\otimes\bm{I})\medskip
\end{equation}
where $\overline{\bm{\mathcal{I}}}_{t|t-1}^{(l)}= \mathbb{E}[\overline{\bm{\nabla}}_t^{(l)}\overline{\bm{\nabla}}_t^{(l)\prime}|\bm{\mathcal{F}}_{t-1}]$. Thus, the normalized score is given by\medskip
\begin{align}
\bm{s}_t^{(l)}&=\left(\bm{\mathcal{I}}_{t|t-1}^{(l)}\right)^{-\alpha}\bm{\nabla}_t^{(l)}\\
&=(\bm{T}^{-\alpha\prime}\otimes \bm{I})\left(\overline{\bm{\mathcal{I}}}_{t|t-1}^{(l)}\right)^{-\alpha}(\bm{T}^{-\alpha}\otimes \bm{I})(\bm{T}\otimes\bm{I})\overline{\bm{\nabla}}_t^{(l)}\\
&=(\bm{T}^{-\alpha\prime}\otimes \bm{I})\left(\overline{\bm{\mathcal{I}}}_{t|t-1}^{(l)}\right)^{-\alpha}(\bm{T}^{-\alpha+1}\otimes \bm{I})    \overline{\bm{\nabla}}_t^{(l)}
\end{align}
We conclude that Equation \eqref{eq:gas_l_partial} can be re-written as\medskip
\begin{equation}
\overline{\bm{l}}_{t+1}= \overline{\bm{c}}^{(l)}+ \overline{\bm{A}}^{(l)} \overline{\bm{s}}_t^{(l)} + \overline{\bm{B}}^{(l)} \overline{\bm{l}}_t
\label{eq:gas_l_transf}
\end{equation}
where 
\begin{align}
\overline{\bm{c}}^{(l)}&=(\bm{T}'\otimes \bm{I})\bm{c}^{(l)},\quad\quad  \overline{\bm{A}}^{(l)}=(\bm{T}'\otimes \bm{I})\bm{A}^{(l)}(\bm{T}'\otimes \bm{I})^{-\alpha}\label{eq:transfL1}\\
 \overline{\bm{B}}^{(l)}&=(\bm{T}'\otimes \bm{I})\bm{B}^{(l)}(\bm{T}'\otimes \bm{I})^{-1},\quad\quad \overline{\bm{s}}_t^{(l)}=\left(\overline{\bm{\mathcal{I}}}_{t|t-1}^{(l)}\right)^{-\alpha}(\bm{T}^{-\alpha+1}\otimes \bm{I})\overline{\bm{\nabla}}_t^{(l)}\label{eq:transfL2}
\end{align}
Similar computations show that Equation \eqref{eq:gas_g} can be written as follows
\begin{equation}
\overline{\bm{g}}_{t+1}=\overline{\bm{c}}^{(g)}+\bm{T}^{-1}\bm{A}^{(g)}\bm{s}_t^{(g)}+\overline{\bm{B}}^{(g)}\overline{\bm{g}}_t\medskip
\end{equation}
where $\overline{\bm{c}}^{(g)}=\bm{T}^{-1}\bm{c}^{(g)}$ and $\overline{\bm{B}}^{(g)}=\bm{T}^{-1}\bm{B}^{(g)}\bm{T}$.\ Using the chain rule as before, we obtain:\medskip
\begin{equation}
\bm{\nabla}_t^{(g)}=\bm{T}^{-1\prime}\overline{\bm{\nabla}}_t^{(g)},\quad\quad \bm{\mathcal{I}}_{t|t-1}^{(g)}=\bm{T}^{-1\prime}\overline{\bm{\mathcal{I}}}_{t|t-1}^{(g)}\bm{T}^{-1}\medskip
\end{equation}
where $\overline{\bm{\nabla}}_t^{(g)}=\left[\frac{\partial\log\mathcal{L}(\bm{y}_t|\bm{\mathcal{F}}_{t-1},\bm{l}_t,\bm{g}_t;\bm{\Theta}_{\bm{\epsilon}})}{\partial\bm{\overline{g}}_t}\right]'$ and $\overline{\bm{\mathcal{I}}}_{t|t-1}^{(g)}= \mathbb{E}[\overline{\bm{\nabla}}_t^{(g)}\overline{\bm{\nabla}}_t^{(g)\prime}|\bm{\mathcal{F}}_{t-1}]$. Therefore:
\begin{align}
\bm{s}_t^{(g)}&=\left(\bm{\mathcal{I}}_{t|t-1}^{(g)}\right)^{-\beta}\bm{\nabla}_t^{(g)}\\
&=\bm{T}^{\beta}\left(\overline{\bm{\mathcal{I}}}_{t|t-1}^{(g)}\right)^{-\beta}(\bm{T}^{\beta\prime})(\bm{T}^{-1\prime})\overline{\bm{\nabla}}_t^{(g)}\\
&=\bm{T}^{\beta}\left(\overline{\bm{\mathcal{I}}}_{t|t-1}^{(g)}\right)^{-\beta}(\bm{T}^{-1+\beta})^{\prime}\overline{\bm{\nabla}}_t^{(g)}
\end{align}
This in turn implies that Equation \eqref{eq:gas_g} can be written as\medskip
\begin{equation}
\overline{\bm{g}}_{t+1}=\overline{\bm{c}}^{(g)}+\overline{\bm{A}}^{(g)}\overline{\bm{s}}_t^{(g)}+\overline{\bm{B}}^{(g)}\overline{\bm{g}}_t\medskip
\label{eq:gas_g_transf}
\end{equation}
where 
\begin{align}
\overline{\bm{c}}^{(g)}&=\bm{T}^{-1}\bm{c}^{(g)},\quad\quad \overline{\bm{A}}^{(g)}=\bm{T}^{-1}\bm{A}^{(g)}\bm{T}^{\beta}  \label{eq:transf_g1_tv}\\
 \overline{\bm{B}}^{(g)}&=\bm{T}^{-1}\bm{B}^{(g)}\bm{T},\quad\quad \overline{\bm{s}}_t^{(g)}=\left(\overline{\bm{\mathcal{I}}}_{t|t-1}^{(g)}\right)^{-\beta}(\bm{T}^{-1+\beta})^{\prime}\overline{\bm{\nabla}}_t^{(g)}. \label{eq:transf_g2_tv}
\end{align}

\subsection{Proof of Theorem \eqref{thm:commut_tv}}
The proof follows exactly the same steps of those of Theorem \eqref{thm:commut}.

\section{Computation of scores and Fisher information matrices}
\label{app:scoresInfo}
In the case of static loadings matrix, the expressions of the score and conditional Fisher information matrix for the Student-$t$ density in Equation \eqref{eq:student-t} have been derived in \cite{creal2014observation}; see their online appendix.\ They are given by:
\begin{align}
\bm{\nabla}_t &= \frac{\nu+n}{\nu-2}\frac{1}{w_t}\bm{\Lambda}'\bm{\Sigma}^{-1}(\bm{y}_t-\bm{\Lambda}\bm{f}_t)\\
\bm{\mathcal{I}}_{t|t-1} &= \frac{\nu}{\nu+n+2}\bm{\Lambda}'\bm{\Sigma}^{-1}\bm{\Lambda},
\end{align} 
where $w_t = 1+\frac{(\bm{y}_t-\bm{\Lambda}\bm{f}_t)'\bm{\Sigma}^{-1}(\bm{y}_t-\bm{\Lambda}\bm{f}_t)}{\nu-2}$.\ In the case of time-varying loadings matrix, the expression of $\bm{\nabla}_t^{(l)}$ is given by:
\begin{equation}
\bm{\nabla}_t^{(l)} = \frac{\nu+n}{\nu-2}\frac{1}{w_t}(\bm{g}_t\otimes \bm{I}_{n\times n})\bm{\Sigma}^{-1}(\bm{y}_t-\bm{\Lambda}_t\bm{g}_t),
\end{equation}
while those of $\bm{\nabla}_t^{(g)}$, $\bm{\mathcal{I}}_{t|t-1}^{(g)}$ are as in the static case, but with $\bm{\Lambda}_t$ in place of $\bm{\Lambda}$.\ The expression of $\bm{\mathcal{I}}_{t|t-1}^{(l)}$ is not needed because  $\bm{\nabla}_t^{(l)} $ is normalized with the identity matrix.
\section{Additional Monte Carlo results}

\subsection{Finite sample properties of the maximum-likelihood and iterative two-step estimators in low dimensions.}
\label{app:sub:MC1}
We report in this section the kernel density estimates of each individual element of $\bm{\Lambda}$, $\bm{\Sigma}$.\ The simulation setting in described in Section \eqref{sub:MC1}.

\begin{figure}[h!]
    \centering 
\begin{subfigure}{0.2\textwidth}
  \includegraphics[width=\linewidth]{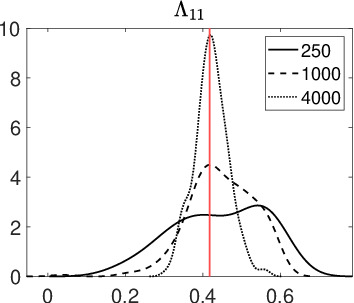}
\end{subfigure}\hfil 
\begin{subfigure}{0.2\textwidth}
  \includegraphics[width=\linewidth]{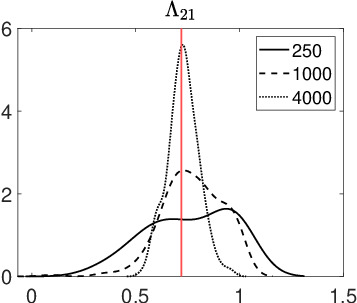}
\end{subfigure}\hfil 
\begin{subfigure}{0.2\textwidth}
  \includegraphics[width=\linewidth]{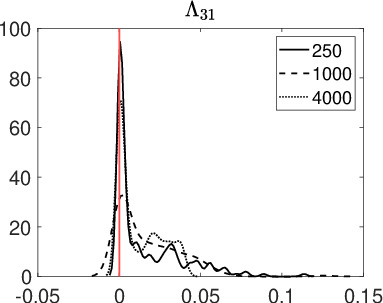}
\end{subfigure}\hfil 
\begin{subfigure}{0.2\textwidth}
  \includegraphics[width=\linewidth]{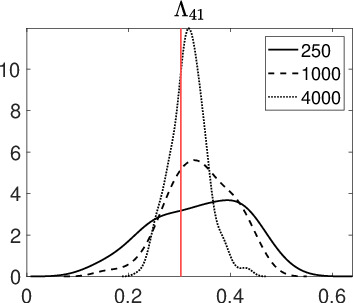}
\end{subfigure}\hfil 
\begin{subfigure}{0.2\textwidth}
  \includegraphics[width=\linewidth]{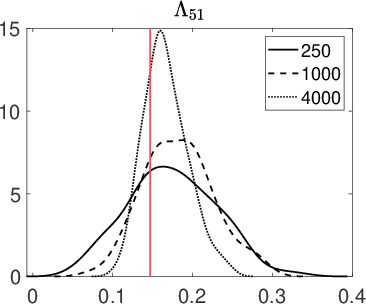}
\end{subfigure}\hfil 

\medskip  

\begin{subfigure}{0.2\textwidth}
  \includegraphics[width=\linewidth]{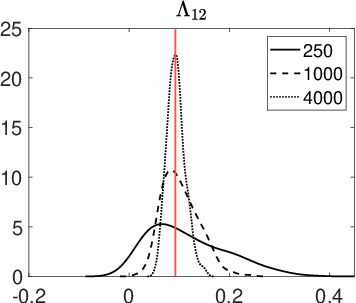}
\end{subfigure}\hfil 
\begin{subfigure}{0.2\textwidth}
  \includegraphics[width=\linewidth]{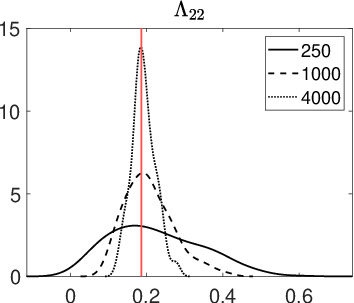}
\end{subfigure}\hfil 
\begin{subfigure}{0.2\textwidth}
  \includegraphics[width=\linewidth]{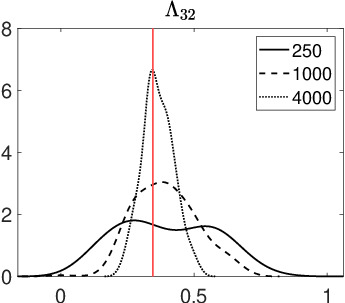}
\end{subfigure}\hfil 
\begin{subfigure}{0.2\textwidth}
  \includegraphics[width=\linewidth]{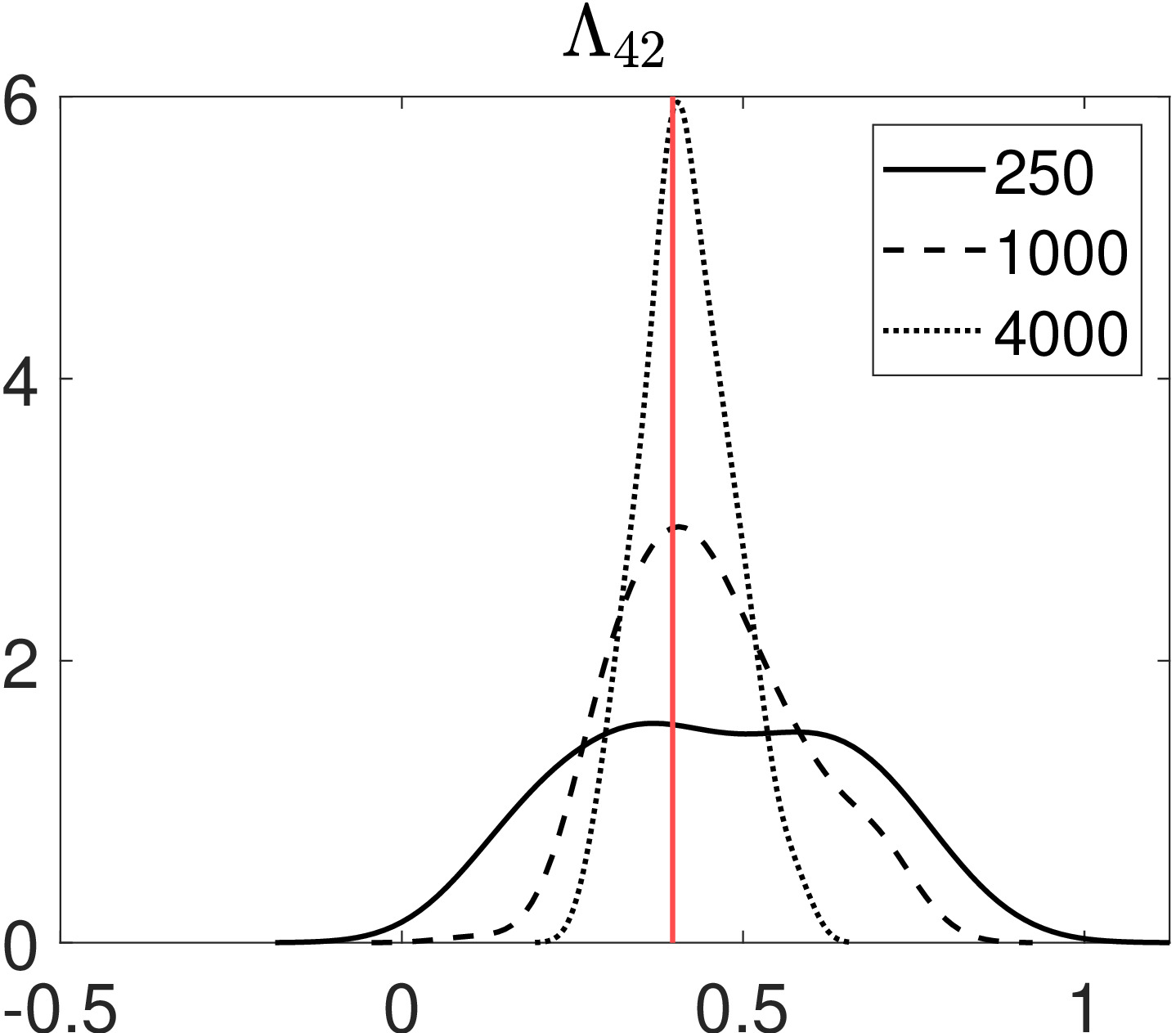}
\end{subfigure}\hfil 
\begin{subfigure}{0.2\textwidth}
  \includegraphics[width=\linewidth]{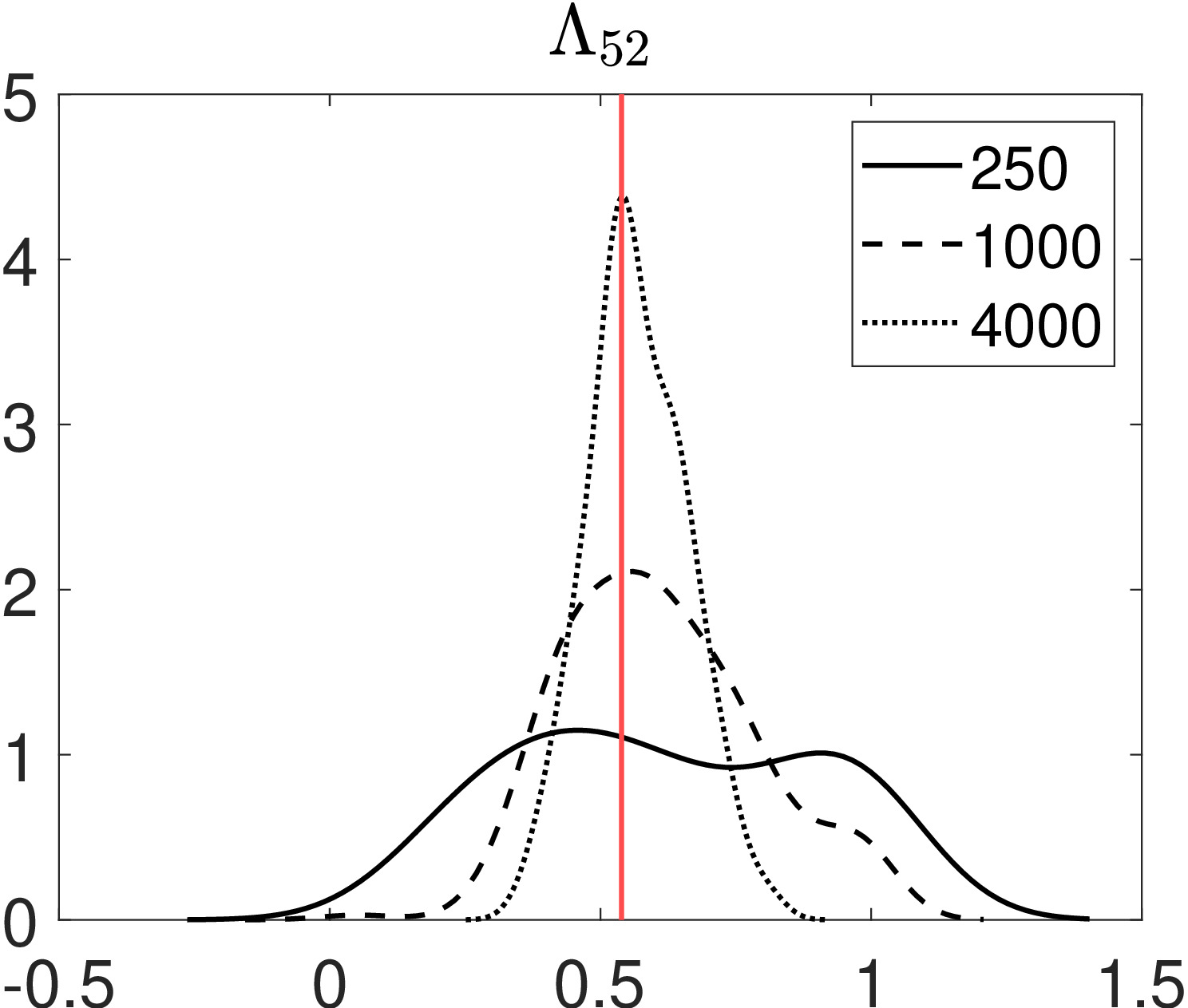}
\end{subfigure}\hfil 
\caption{Kernel density estimates of the maximum likelihood estimates of each individual element of $\bm{\Lambda}$. The estimates are obtained based on $N=250$ replications of a score-driven factor model with sample size $T=250,1000,4000$.\ The vertical red line denotes the true parameter value. }
\end{figure}

\begin{figure}[h!]
    \centering 
\begin{subfigure}{0.3\textwidth}
  \includegraphics[width=\linewidth]{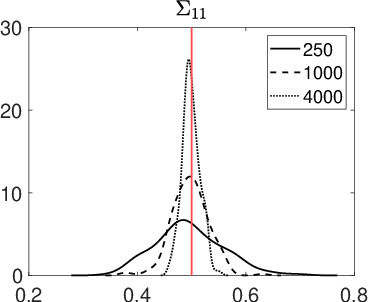}
\end{subfigure}\hfil 
\begin{subfigure}{0.3\textwidth}
  \includegraphics[width=\linewidth]{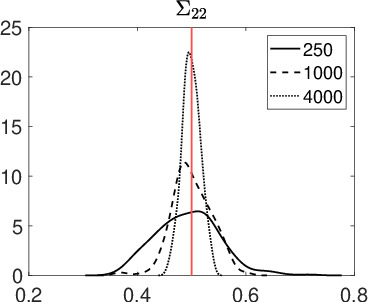}
\end{subfigure}\hfil 
\begin{subfigure}{0.3\textwidth}
  \includegraphics[width=\linewidth]{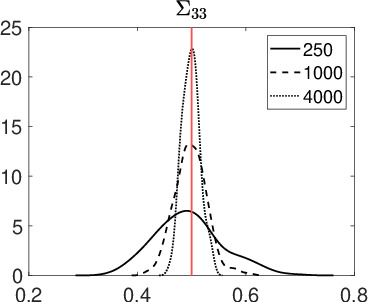}

\medskip  
  
\end{subfigure}\hfil 
\begin{subfigure}{0.3\textwidth}
  \includegraphics[width=\linewidth]{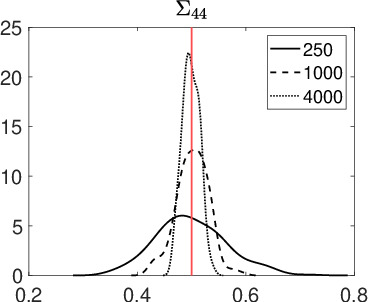}
\end{subfigure}\hfil 
\begin{subfigure}{0.3\textwidth}
  \includegraphics[width=\linewidth]{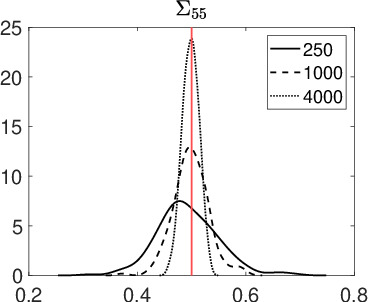}
\end{subfigure}\hfil 
\caption{Kernel density estimates of the maximum likelihood estimates of each individual element of $\bm{\Sigma}$. The estimates are obtained based on $N=250$ replications of a score-driven factor model with sample size $T=250,1000,4000$.\ The vertical red line denotes the true parameter value. }
\end{figure}

\subsection{Finite sample properties of the maximum-likelihood estimator in the time-varying loadings case.}
\label{app:sub:MC4}
We report in this section the kernel density estimates of each individual element of $\bm{c}^{(l)}$, $\bm{A}^{(l)}$, $\bm{B}^{(l)}$, $\bm{\Sigma}$ in the case of time-varying loadings.\ The simulation setting in described in Section \eqref{sub:MC4}.

\begin{figure}[h!]
    \centering 
\begin{subfigure}{0.2\textwidth}
  \includegraphics[width=\linewidth]{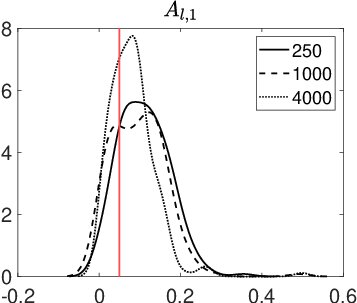}
\end{subfigure}\hfil 
\begin{subfigure}{0.2\textwidth}
  \includegraphics[width=\linewidth]{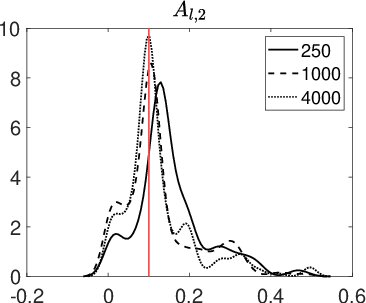}
\end{subfigure}\hfil 
\begin{subfigure}{0.2\textwidth}
  \includegraphics[width=\linewidth]{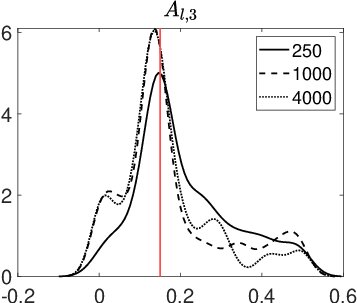}
\end{subfigure}\hfil 
\begin{subfigure}{0.2\textwidth}
  \includegraphics[width=\linewidth]{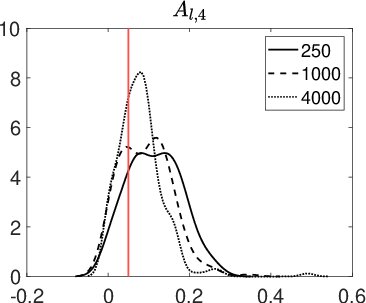}
\end{subfigure}\hfil 
\begin{subfigure}{0.2\textwidth}
  \includegraphics[width=\linewidth]{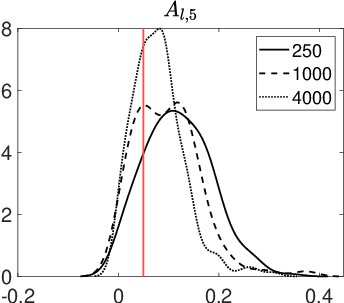}
\end{subfigure}\hfil 

\medskip  

\begin{subfigure}{0.2\textwidth}
  \includegraphics[width=\linewidth]{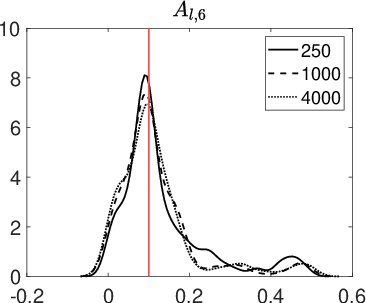}
\end{subfigure}\hfil 
\begin{subfigure}{0.2\textwidth}
  \includegraphics[width=\linewidth]{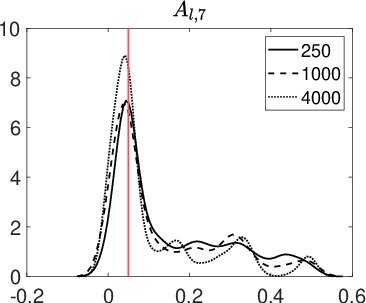}
\end{subfigure}\hfil 
\begin{subfigure}{0.2\textwidth}
  \includegraphics[width=\linewidth]{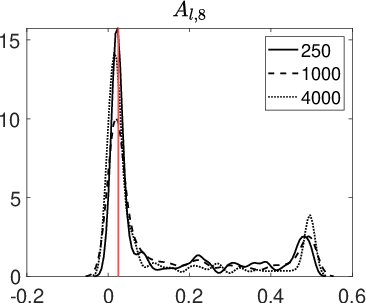}
\end{subfigure}\hfil 
\begin{subfigure}{0.2\textwidth}
  \includegraphics[width=\linewidth]{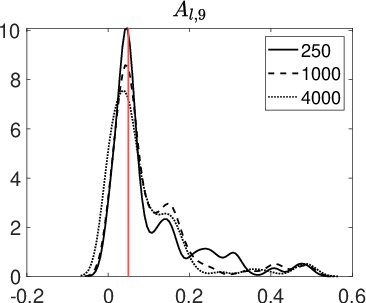}
\end{subfigure}\hfil 
\begin{subfigure}{0.2\textwidth}
  \includegraphics[width=\linewidth]{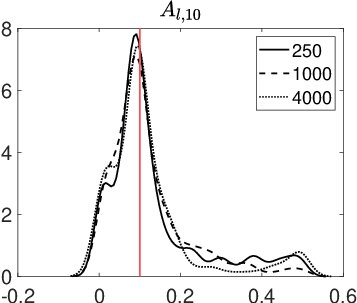}
\end{subfigure}\hfil 
\caption{Kernel density estimates of the maximum likelihood estimates of each diagonal element of $\bm{A}^{(l)}$. The estimates are obtained based on $N=250$ replications of a score-driven factor model with time-varying loadings with sample size $T=250,1000,4000$.\ The vertical red line denotes the true parameter value. }
\end{figure}


\begin{figure}[h!]
    \centering 
\begin{subfigure}{0.2\textwidth}
  \includegraphics[width=\linewidth]{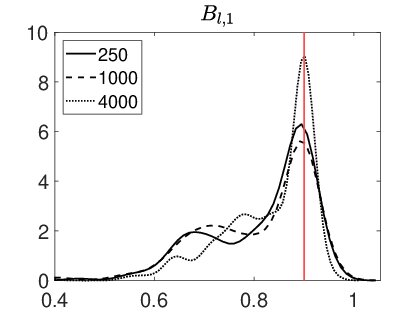}
\end{subfigure}\hfil 
\begin{subfigure}{0.2\textwidth}
  \includegraphics[width=\linewidth]{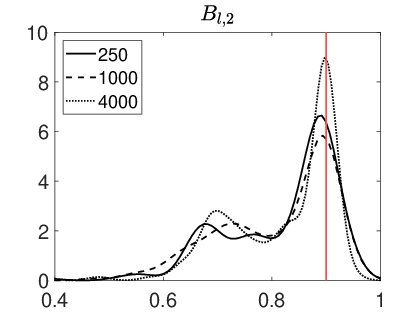}
\end{subfigure}\hfil 
\begin{subfigure}{0.2\textwidth}
  \includegraphics[width=\linewidth]{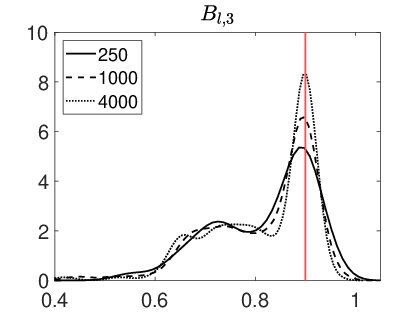}
\end{subfigure}\hfil 
\begin{subfigure}{0.2\textwidth}
  \includegraphics[width=\linewidth]{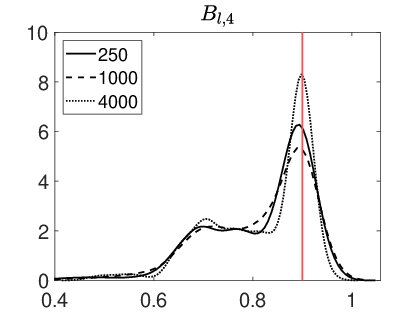}
\end{subfigure}\hfil 
\begin{subfigure}{0.2\textwidth}
  \includegraphics[width=\linewidth]{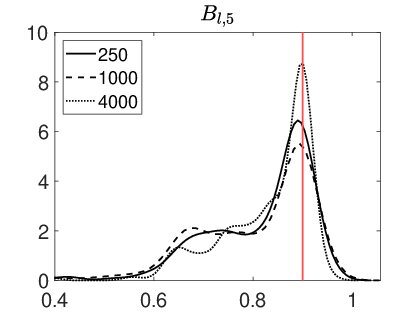}
\end{subfigure}\hfil 

\medskip  

\begin{subfigure}{0.2\textwidth}
  \includegraphics[width=\linewidth]{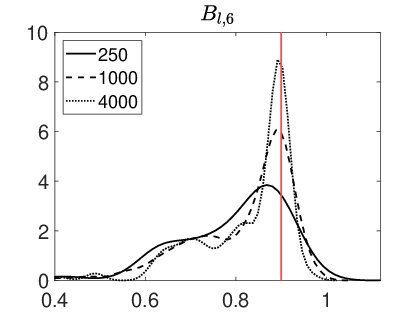}
\end{subfigure}\hfil 
\begin{subfigure}{0.2\textwidth}
  \includegraphics[width=\linewidth]{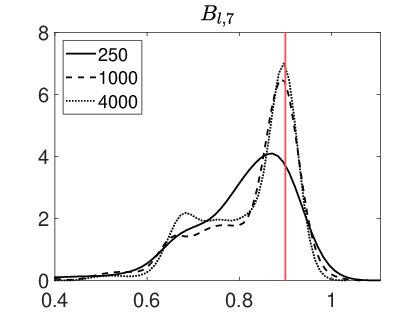}
\end{subfigure}\hfil 
\begin{subfigure}{0.2\textwidth}
  \includegraphics[width=\linewidth]{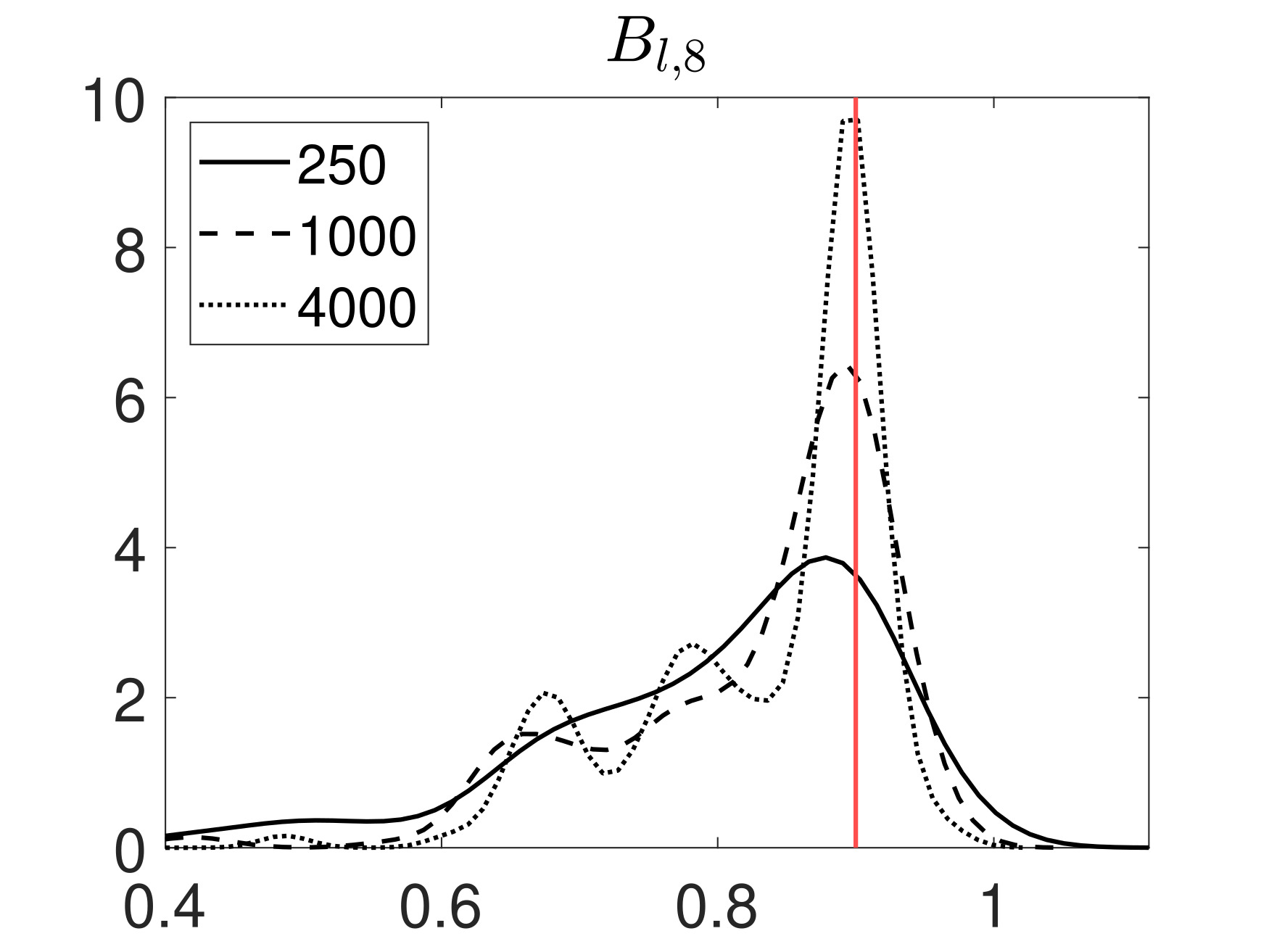}
\end{subfigure}\hfil 
\begin{subfigure}{0.2\textwidth}
  \includegraphics[width=\linewidth]{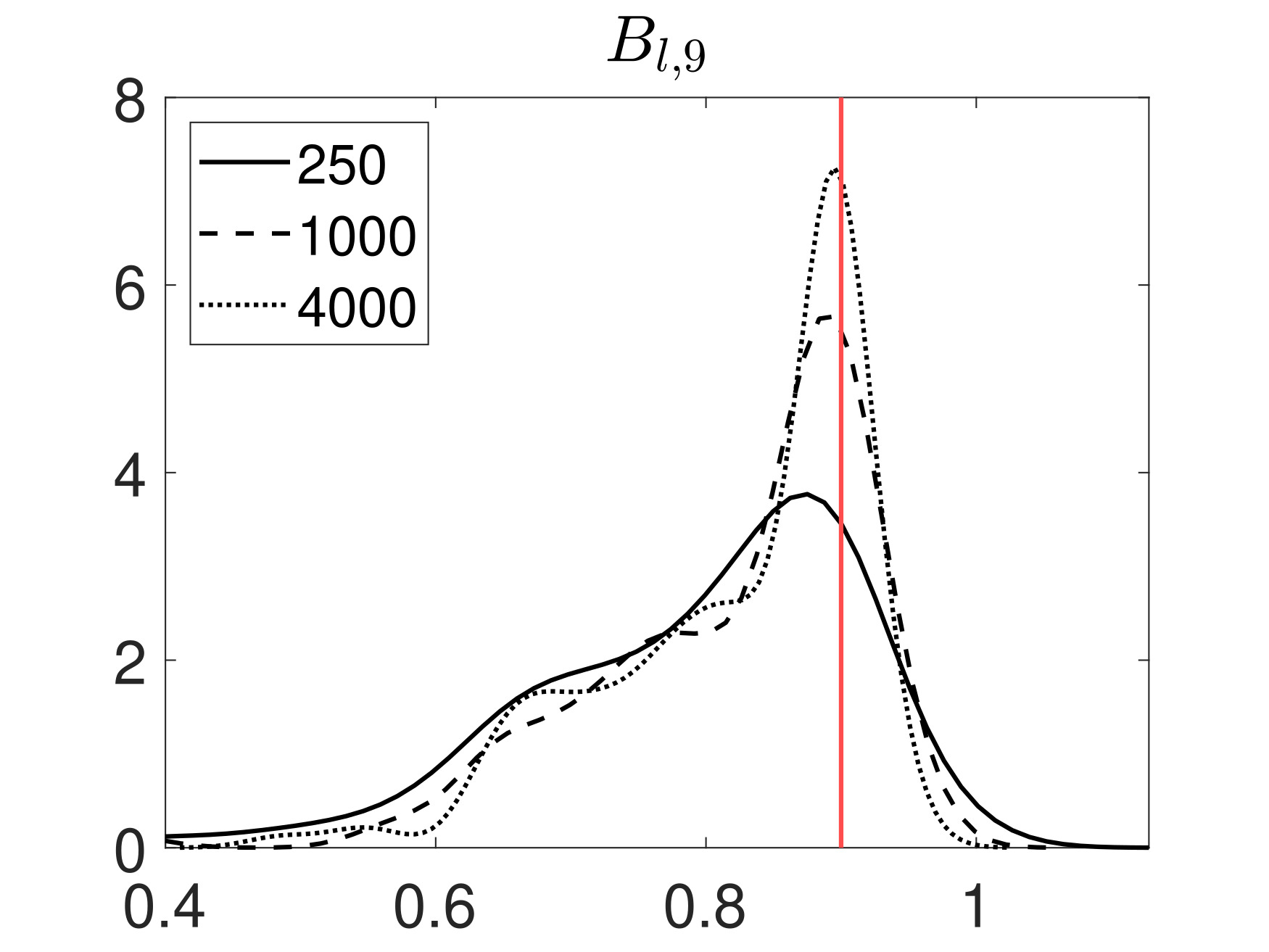}
\end{subfigure}\hfil 
\begin{subfigure}{0.2\textwidth}
  \includegraphics[width=\linewidth]{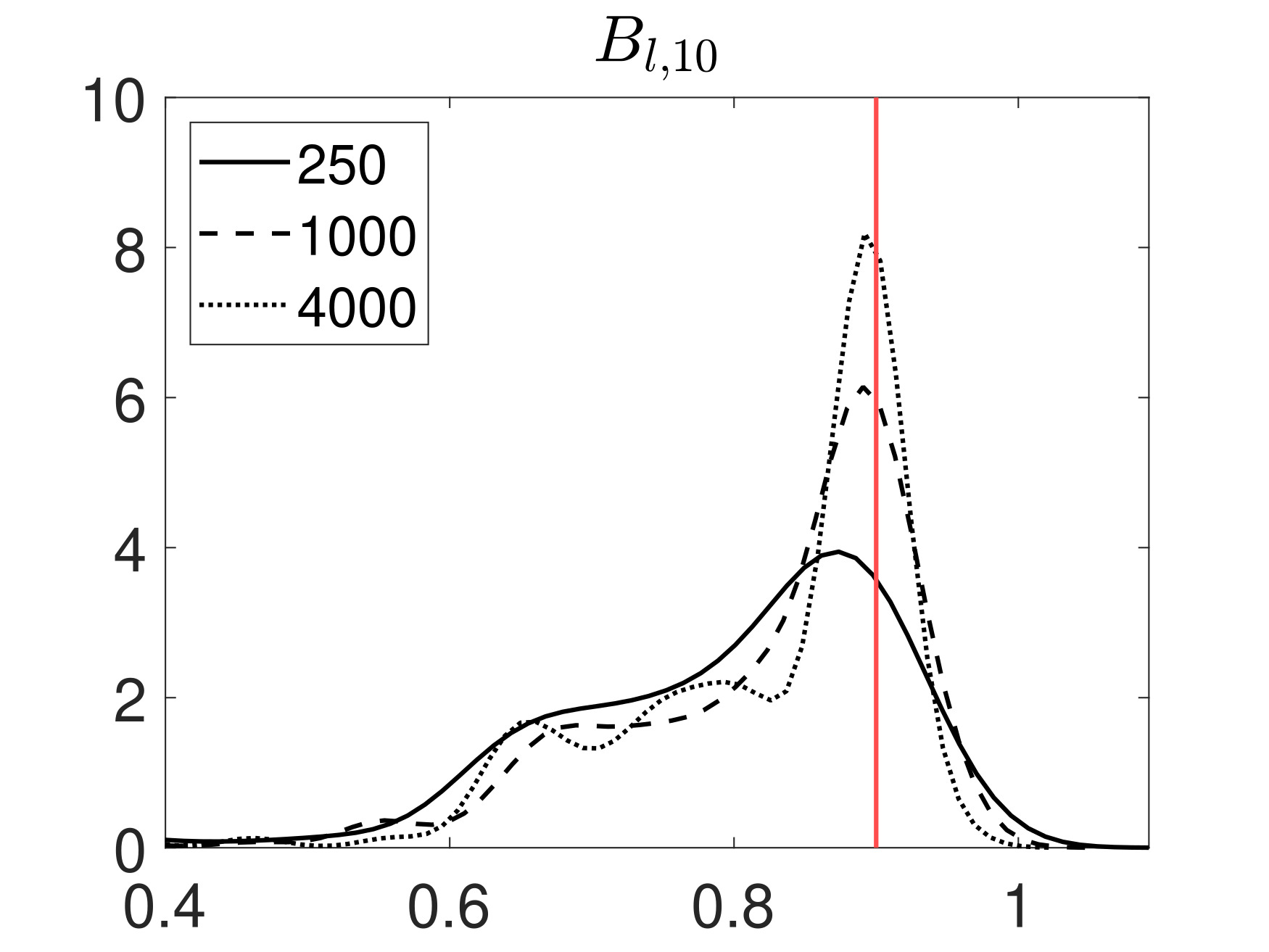}
\end{subfigure}\hfil 
\caption{Kernel density estimates of the maximum likelihood estimates of each diagonal element of $\bm{B}^{(l)}$. The estimates are obtained based on $N=250$ replications of a score-driven factor model with time-varying loadings with sample size $T=250,1000,4000$.\ The vertical red line denotes the true parameter value. }
\end{figure}

\begin{figure}[h!]
    \centering 
\begin{subfigure}{0.4\textwidth}
  \includegraphics[width=\linewidth]{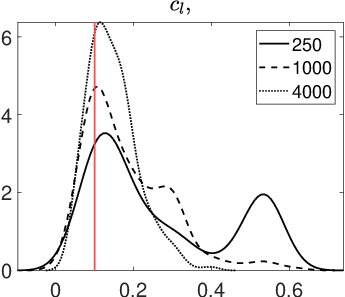}
\end{subfigure}\hfil 
\begin{subfigure}{0.4\textwidth}
  \includegraphics[width=\linewidth]{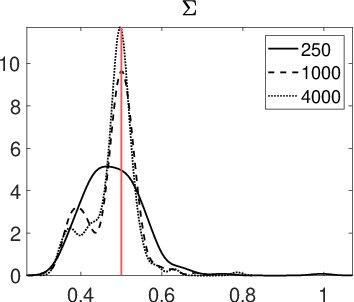}
\end{subfigure}\hfil 
\caption{Kernel density estimates of the maximum likelihood estimates of $\bm{c}^{(l)}$ and $\bm{\Sigma}$.\ In the case of $\bm{\Sigma}$, we report the average of the density estimates for each diagonal element.\  The estimates are obtained based on $N=250$ replications of a score-driven factor model with time-varying loadings with sample size $T=250,1000,4000$.\ The vertical red line denotes the true parameter value.}
\end{figure}

\clearpage
\renewcommand{\thetable}{D\arabic{table}}
\setcounter{table}{0}
\section{Dataset for the application to daily returns of S\&P500 constituents}\label{app:data} 
\begin{table} [h!] 
\caption{Ticker symbols and industrial classification} \label{table:D1}
\scalebox{0.8}{
\begin{tabular}{l|c}
\hline
Tickers&Industry\\
\hline
\hline
AES,ATI,DD,FLR,IP,LPX,PG & Basic Industries\\
\hline
A,AA,BA,CAT,F,GD,HON,NOC & Capital Goods\\
\hline
CL,CPB,EL,GIS,KO,MO,PEP,SYY & Consumer Non-Durables\\
\hline
ANF,BXP,DIS,EQR,HD,JCP,MCD,NLY,RCL,TGT,TV,WMT,WSM,WY & Consumer Services\\
\hline
CNX,CVX,GE,HAL,MUR,OXY,SLB,SU,XOM & Energy\\
\hline
AIG,AXP,BAC,C,COF,GS,HIG,JPM,KEY,MCO,MMC,MS,MTB,PNC,USB,WFC & Financials\\
\hline
ABT,BAX,BMY,CI,JNJ,LLY,MDT,MMM,MRK,PFE,THC & Health Care\\
\hline
AEP,DUK,EXC,SO,VZ,WMB & Public Utilities\\
\hline
DOV,HPQ,IBM,TSM & Technology\\
\hline
FDX,LUV,NSC,UPS & Transportation\\
\hline
\end{tabular}}
\end{table}

\end{document}